\documentclass[journal,twoside,web]{ieeecolor}
\usepackage{graphicx}
\usepackage{balance} %
\usepackage{float}
\usepackage{rotating}
\usepackage{subcaption}
\usepackage{psfrag}
\usepackage{amsmath,amsfonts,bbm,nicefrac,mathtools,lipsum}
\usepackage{multirow}
\usepackage{latexsym}
\usepackage{caption}
\usepackage{colortbl}
\usepackage[super]{nth}
\usepackage{fancyhdr}
\usepackage{epstopdf}
\usepackage{booktabs,tabularx}
\usepackage{makecell}
\usepackage{dirtytalk}
\usepackage{hyperref}
\usepackage{siunitx}
\usepackage{academicons}
\usepackage{bbm}
\usepackage{nccmath} 

\newcommand{\twnote}[1]%
    {\textcolor{cyan}{\textbf{TW: #1}}}

\newenvironment{customproof}[1][Proof]{%
  \par\noindent\textbf{#1.}\quad%
}{\hfill$\blacksquare$\\}

\newenvironment{appendixsec}[1][Proof]{%
  \par\noindent\textbf{#1.}\quad%
}

\newcommand{\tacinitial}[1]{{\leavevmode\color{black}#1}}

\makeatletter
\usepackage{xspace}
\DeclareRobustCommand\onedot{\futurelet\@let@token\@onedot}
\def\@onedot{\ifx\@let@token.\else.\null\fi\xspace}
\def\eg{e.g\onedot} 
\def\ie{i.e\onedot}

\makeatother

\usepackage{soul}
\usepackage[ruled]{algorithm2e}
\usepackage[capitalize]{cleveref}

\usepackage[acronyms, shortcuts]{glossaries}
\glsdisablehyper

\newacronym{ioc}{IOC}{inverse optimal control}
\newacronym{lqr}{LQR}{linear-quadratic regulator}
\newacronym{kkt}{KKT}{Karush–Kuhn–Tucker}
\newacronym{irl}{IRL}{inverse reinforcement learning}
\newacronym{mle}{MLE}{maximum likelihood estimation}
\newacronym{sem}{SEM}{standard error of the mean}
\newacronym{mpgp}{MPGP}{model-predictive game play}
\newacronym[longplural=open-loop Nash equilibria,plural=OLNE]{olne}{OLNE}{open-loop Nash equilibrium}
\newacronym{licq}{LICQ}{linear independence constraint qualification}
\newacronym{mpc}{MPC}{model-predictive control}
\newacronym{mpec}{MPEC}{mathematical program with equilibrium constraints}
\newacronym[longplural={partially observable Markov decision processes}]{pomdp}{POMDP}{partially observable Markov decision process}
\newacronym{nep}{NEP}{Nash equilibrium problem}
\newacronym{gnep}{GNEP}{generalized Nash equilibrium problem}
\newacronym{lq}{LQ}{linear-quadratic}
\newacronym{gne}{GNE}{generalized Nash equilibrium}
\newacronym{svo}{SVO}{social value orientation}
\newacronym{ukf}{UKF}{unscented Kalman filter}
\newacronym{ibr}{IBR}{iterated best response}
\newacronym{awgn}{AWGN}{additive white Gaussian noise}
\newacronym{iqr}{IQR}{interquartile range}
\newacronym{mcp}{MCP}{Mixed Complementarity Problem}
\newacronym{ift}{IFT}{implicit function theorem}
\newacronym{nn}{NN}{neural network}
\newacronym{rnn}{RNN}{recurrent neural network}
\newacronym{lstm}{LSTM}{long short-term memory}
\newacronym{gru}{GRU}{gated recurrent unit}
\newacronym{bilstm}{BiLSTM}{bidirectional long short-term memory}
\newacronym{hj}{HJ}{Hamilton–Jacobi}
\newacronym{admm}{ADMM}{alternating direction method of multipliers}
\newacronym{dp}{DP}{dynamic programming}
\newacronym{iss}{ISS}{input-to-state stable}
\newacronym{las}{LAS}{locally asymptotically stable}
\newcommand{\norm}[1]{\| #1 \|}
\newcommand{\normf}[1]{\|#1\|_{F}}
\newcommand{\inv}[1]{#1^{-1}}
\newcommand{\maxsv}[1]{\sigma_{\mathrm{max}}(#1)}
\newcommand{\minsv}[1]{\sigma_{\mathrm{min}}(#1)}
\newcommand{\given}{\mid}
\newcommand{\mbb}{\mathbb}

\newcommand{\R}{\mbb{R}}

\newcommand{\numplayers}{N}

\newcommand{\states}{x}

\newcommand{\control}{u}

\newcommand{\horizon}{T}
\newcommand{\cost}{J}

\newcommand{\St}{S_t}
\newcommand{\Yt}{Y_t}
\newcommand{\Pt}{P_t}
\newcommand{\Mt}{M_t}

\newcommand{\Fht}{\hat{F}_t}

\newcommand{\Pht}{\hat{P}_t}

\newcommand{\thetat}{\Tilde{\theta}^{*}_{i,j}}
\newcommand{\vect}{\mathrm{vec}}
\newcommand{\fhat}{\hat{h}}

\newcommand{\DynRiccati}{\tau}

\newcommand{\StateRiccatiAll}{\Tilde{Z}}
\newcommand{\StateRiccatiReg}{\hat{\Tilde{Z}}^i}
\newcommand{\StateRiccatiAllReg}{\hat{\Tilde{Z}}}
\newcommand{\BallRiccati}{\rho}
\newcommand{\RiccatiTransient}{\beta}
\newcommand{\MinSingularValue}{\nu}
\newcommand{\RegRiccatiAutonomous}{\phi}
\newcommand{\RegRiccatiDisturb}{O}

\newcommand{\ValueFunc}{V}
\newcommand{\half}{\frac{1}{2}}
\newcommand{\StateStageCost}[1]{\half (\states_{#1}^{\top}Q_{#1}^i + 2q_{#1}^{i\top})\states_{#1}}
\newcommand{\ControlStageCost}[1]{\half \sum_{j = 1}^{\numplayers}(\control_{#1}^{j\top}R_{#1}^{ij} + 2r_{#1}^{ij\top})\control_{#1}^j}
\newcommand{\At}{A_t}
\newcommand{\Bjt}{B^j_t}
\newcommand{\ExpandNextStateStageCost}[1]{\half [(\At \states_t + \sum_{j \in [N]}\Bjt \control_t^j)^{\top}Q_{t+1}^i + 2q_{t+1}^{i\top}](\At \states_t + \sum_{j \in [N]}\Bjt \control_t^j)}

\newcommand{\example}[1]%
{
\vspace{0.15cm}
\noindent \textit{\textbf{Running example:} #1}
\vspace{0.15cm}
}
\newcommand{\ith}{$i^\mathrm{th}$~}
\newcommand{\jth}{$j^\mathrm{th}$~}
\newtheorem{theorem}{Theorem}
\newtheorem{remark}{Remark}
\newtheorem{assumption}{Assumption}
\newtheorem{definition}{Definition}

\newtheorem{lemma}{Lemma}
\newtheorem{corollary}{Corollary}

\newcommand{\Z}{\mathbb{Z}}
\newcommand{\N}{\mathbb{N}}

\newcommand{\revised}[1]{{\leavevmode\color{black}#1}}

\graphicspath{{./Images/}}

\usepackage{generic}
\usepackage{cite}
\usepackage{amsmath,amssymb,amsfonts}
\usepackage{graphicx}

\usepackage{textcomp}
\def\BibTeX{{\rm B\kern-.05em{\sc i\kern-.025em b}\kern-.08em
    T\kern-.1667em\lower.7ex\hbox{E}\kern-.125emX}}
\begin{document}
\title{
Approximate Feedback Nash Equilibria with Sparse Inter-Agent Dependencies
}


\author{Xinjie Liu, \IEEEmembership{Graduate Student Member, IEEE}, Jingqi Li, \IEEEmembership{Graduate Student Member, IEEE}, Filippos Fotiadis, \IEEEmembership{Member, IEEE}, Mustafa O. Karabag, Jesse Milzman, \IEEEmembership{Member, IEEE}, David Fridovich-Keil, \IEEEmembership{Member, IEEE}, Ufuk Topcu, \IEEEmembership{Fellow, IEEE}
\thanks{
\tacinitial{
This work was supported in part by the Office of Naval Research (ONR) under Grants ONR N00014-24-1-2797 and ONR N00014-22-1-2703, and in part by the National Science Foundation (NSF) under Grants No. 2409535 and No. 2336840, and in part by the Army Research Laboratory and was accomplished under Cooperative Agreement Number W911NF-23-2-0011.}
}
\thanks{
\tacinitial{
X. Liu, F. Fotiadis, M. O. Karabag, D. Fridovich-Keil, and U. Topcu are with the Oden Institute for Computational Engineering and Sciences, The University of Texas at Austin, TX, 78712, USA (emails: \texttt{\{xinjie-liu,ffotiadis,karabag,dfk,utopcu\}@utexas.edu}). 
J. Li is with the Department of Electrical Engineering and Computer Sciences, University of California, Berkeley, CA, 94720-2284, USA (email: \texttt{jingqili@berkeley.edu}). 
J. Milzman is with the DEVCOM Army Research Laboratory, MD, 20783-1138, USA (email: \texttt{jesse.m.milzman.civ@army.mil}).} 
}}

\maketitle

\begin{abstract}
Feedback Nash equilibrium strategies in multi-agent dynamic games require availability of all players' state information to compute control actions. 
However, in real-world scenarios, sensing and communication limitations between agents make full state feedback expensive or impractical, and such strategies can become fragile when state information from other agents is inaccurate. 
To this end, we propose a regularized dynamic programming approach for finding \emph{sparse} feedback policies that selectively depend on the states of a subset of agents in dynamic games. 
The proposed approach solves convex adaptive group Lasso problems to compute sparse policies approximating Nash equilibrium solutions. 
We prove the regularized solutions' asymptotic convergence to a neighborhood of Nash equilibrium policies in \ac{lq} games. 
Further, we extend the proposed approach to general non-\ac{lq} games via an iterative algorithm. 
Simulation results in multi-robot interaction scenarios show that the proposed approach effectively computes feedback policies with varying sparsity levels. 
When agents have noisy observations of other agents' states, simulation results indicate that the proposed regularized policies consistently achieve lower costs than standard Nash equilibrium policies by up to 77\% for all interacting agents whose costs are coupled with other agents' states.\footnote{\href{https://xinjie-liu.github.io/projects/sparse-games}{Project website: \color{magenta}https://xinjie-liu.github.io/projects/sparse-games}}
\end{abstract}


\section{Introduction}\label{sec:intro}

Dynamic game theory models the strategic decisions of multiple interacting agents over time. 
In such games, it is common to identify solutions at which players execute \emph{full state feedback} strategies that depend on \emph{all} players' states. 
For example, in multi-robot formations, each robot typically plans its actions based on the states of other robots. 
However, obtaining full state information is often expensive or impractical due to sensing and communication limitations. 
Worse, \say{dense} strategies that require access to many other agents' states can be brittle when such state information is inaccurate, \eg, in the presence of uncertainties. 
Consequently, it is desirable for agents to find strategies that selectively depend on the states of a subset of agents while still approximating equilibrium behavior.

We contribute an algorithm for finding \emph{sparse} feedback policies that depend on fewer influential agents' states in dynamic games while approximating Nash equilibrium strategies.  
More specifically, we propose a regularized \ac{dp} scheme~\cite{bacsar1998dynamic,bertsekas2012dynamic} that approximately solves \acf{lq} dynamic games, which are an important extension of the \ac{lqr} problem~\cite{kalman1960contributions} to multi-agent interaction settings. 
The proposed approach solves a \emph{convex} adaptive group Lasso 
regularization problem~\cite{wang2008note} to encode structured sparsity within each \ac{dp} iteration. 
A user can choose a desired sparsity level based on the available sensing or communication resources. 
We also employ an iterative linear-quadratic approximation technique~\cite{fridovich2020efficient,laine2021computation} to extend the proposed approach to general non-\ac{lq} games. 

To establish theoretical properties of the proposed approach, we first derive an upper bound on the deviation of the regularized policy matrix computed by the proposed approach from the Nash equilibrium in a single stage of an \ac{lq} game. 
Then, we prove the asymptotic convergence of the regularized \ac{dp} recursion to a neighborhood of the Nash equilibrium across all stages of the game.

We test the proposed approach in simulation and support the following key claims: First, the proposed approach yields more sparse feedback strategies than standard Nash equilibrium solutions and thereby requires less communication and sensing resources to execute. Second, the proposed approach yields different levels of sparsity as the user varies the regularization strength.  
Third, we empirically show that the proposed approach can be readily extended to general non-\ac{lq} games via an iterative scheme. 
Fourth, for all interacting agents whose costs are coupled with other agents’ states, the regularized strategies show improved robustness and task performance in comparison to standard Nash equilibria in cases where agents only have inaccurate knowledge of other agents, \eg, noisy estimates of other agents' states.

\section{Related Works}\label{sec:related-works}

\subsection{Noncooperative Games}\label{sec:noncooperative-games}

Over the years, a growing body of literature has explored the application of dynamic noncooperative games~\cite{bacsar1998dynamic,isaacs1999differential} in designing autonomous systems~\cite{schwarting2019social,fridovich2020efficient,fisac2019hierarchical,9329208,mehar2023maximum,yu2023active,peters2023ijrr,liu2024auto,li2023cost,hu2024think,gupta2024second}. 
Among different types of games, we focus on general-sum Nash equilibrium problems, where players can have partially competing objectives and make decisions simultaneously. 
In Nash games with different information patterns,  feedback and open-loop Nash equilibrium are the most common solution types.
Open-loop Nash equilibria ignore the dynamic, multi-stage nature of the problem by assuming that players choose their entire decision sequences, \ie, trajectories, at once~\cite{cleac2022algames,zhu2023sequential,mcp_ref,liu2023learning,peters2024ral}. 
In contrast, a feedback Nash equilibrium is a set of~control \emph{policies} that map game states to players' controls; they are more expressive than open-loop solutions but strictly more challenging to compute. 
Various special cases have been studied, \eg, in~\cite{7456255,8848414,kossioris2008feedback}, among others. 
For general feedback games with nonconvex costs and constraints, only approximate solutions exist, \eg, in~\cite{fridovich2020efficient,laine2021computation}.
We also note that both open-loop and feedback \ac{lq} games have well-studied solutions~\cite{bacsar1998dynamic,lyapunov_iteration}.

This work focuses on feedback solutions to \ac{lq} dynamic games, where we encourage sparsity structure in the resulting strategies via a regularized \ac{dp} procedure. We also extend the approach to non-\ac{lq} games via an iterative linear-quadratic approximation technique~\cite{fridovich2020efficient}.

\subsection{Structured Controller Design}

The literature on encoding sparsity structures into control strategies primarily focuses on single-agent optimal control settings~\cite{kalman1960contributions} and can generally be categorized into two groups. 
The first category~\cite{wenk1980parameter,lin2011augmented,maartensson2012scalable,lamperski2012optimal,fardad2009optimal,furieri2020learning} requires a predefined sparsity structure, using this as a constraint to compute controls. 
For instance, each controller may have access only to a subset of the sensing information. 
More relevant to our work, studies in the second category propose methods to encourage sparsity in classical \ac{lqr} problems without requiring a predefined sparsity structure \cite{lin2013design,dorfler2014sparsity,wytock2013fast,park2020structured}. These approaches introduce sparsity regularizers into the \ac{lqr} cost, resulting in nonconvex policy optimization problems. 
\tacinitial{Prior works further encourage sparse policies in cooperative Markov games with finite state-action spaces \cite{karabag2022planning} and noncooperative \ac{lq} games \cite{lian2017game}. 
Although \cite{lian2017game} also studies noncooperative games, it does not consider group sparsity that explicitly reduces inter-agent dependencies; rather, it requires the number of nonzero entries in policy matrices to be specified a priori and as constraints, which does not offer flexibility in balancing between game costs and sparsity level. 
}

Research on finding sparse policies in noncooperative games remains relatively scarce, despite the fact that sparse strategies are especially desirable in reducing inter-agent communication and sensing requirements. 
To address this gap, our approach encourages \emph{group} sparsity in the computed control policies, where each group corresponds to the influence of one agent’s state on another agent’s control action. 
Given the convergence challenges associated with direct policy optimization in general-sum games, we adopt a modified \ac{dp} approach. At each stage, we solve a \emph{convex} adaptive group Lasso problem \cite{wang2008note,yuan2006model}, an extension of the Lasso problem~\cite{tibshirani1996regression} to variables in pre-specified groups. 
This modified \ac{dp} scheme retains convexity and enables formal convergence analysis and extension to more general non-\ac{lq} game settings. 

\tacinitial{
A preliminary version of this work has been accepted for presentation at the International Conference on Autonomous Agents and Multiagent Systems, 2025\footnote{We include the accepted version in supplementary materials for reference.}. This manuscript additionally adds theoretical results, proofs, more experimental results, a discussion of related works, and a more detailed presentation of the approach.}

\section{Preliminaries}

We study noncooperative dynamic games played by $\numplayers$ players in discrete time~$t \in \{1, \dots, \horizon\} \equiv [\horizon]$. 
\tacinitial{
We first study \ac{lq} games in \cref{sec:lq-case}, and then extend to more general non-\ac{lq} games in \cref{sec:non-lq-case}. We define \ac{lq} games as the following:
\begin{definition}\label{def:lq-game}
An \numplayers-player, general-sum, discrete-time dynamic game is a \acf{lq} game if each player $i \in \{1, \dots, \numplayers\} \equiv [\numplayers]$ seeks to optimize an individual quadratic cost function:
\begin{multline}
\label{eq:lq-game-objective}
    \cost^i = \frac{1}{2} \sum_{t=1}^T\Big((\states_t^{\top}Q_t^i + 2q_t^{i\top})\states_t + \sum_{j = 1}^{\numplayers}(\control_t^{j\top}R_t^{ij} + 2r_t^{ij\top})\control_t^j
    \Big) \\ + \frac{1}{2} (\states_{T+1}^{\top}Q_{T+1}^i + 2q_{T+1}^{i\top})\states_{T+1},
\end{multline}
and players follow a joint linear dynamical system:
\begin{equation}\label{eq:lq-game-dynamics}
    \states_{t+1} = A_t\states_t + \sum_{i = 1}^\numplayers B_t^i\control_t^i,
\end{equation}
with~$x_t^i \in \R^{m^i}, u_t^i \in \R^{n^i}$, and~$Q_t^i \succeq 0, R^{ii}_t \succ 0, \forall i \in [\numplayers]$. 
We let~$m = \sum_{i \in [\numplayers]}m^i$ and~$n = \sum_{i \in [\numplayers]}n^i$. The initial state of the game~$\states_1$ is a given a priori.
\end{definition}

\noindent\textbf{Notation.} $\states \in \R^{m}$ and~$\control \in \R^{n}$ denote players' states and controls. Superscripts denote players' indices and subscripts are discrete time steps, \eg,~$\control_t^i \in \R^{n^i}$ denotes player~$i$'s control at time step~$t$. 
Throughout the manuscript, the absence of player and time indices without additional definition denotes concatenation, \ie,~$\states_{t} := [\states^{1\top}_t, \states^{2\top}_t, \hdots, \states^{\numplayers\top}_t]^\top$ and $\states^i := [\states^{i\top}_1, \states^{i\top}_2, \hdots, \states^{i\top}_\horizon]^\top$.

In more general non-\ac{lq} games discussed in \cref{sec:non-lq-case}, players follow a joint  dynamical system:
\begin{equation}
    \states_{t+1} = f_t(\states_t, \control^1_t, \hdots, \control^\numplayers_t),
\end{equation}
while optimizing individual costs~$\cost^i$:
\begin{equation}\label{eq:general-game-cost}
    \min_{\control^i} \underbrace{\sum_{t=1}^T g_t(\states_t, \control^1_t, \hdots, \control^\numplayers_t) + g_{T+1}(\states_{T+1})}_{\cost^i(\control_1, \dots, \control_{\horizon})},
\end{equation}
consisting of additive stage costs~$g_t, \forall t \in [T]$ and a terminal cost~$g_{T+1}$. 
The cost functions $\cost^i$ can be nonquadratic and the dynamics $f_t$ can be nonlinear functions. 
}


We focus on the solution concept of feedback Nash equilibrium~\cite[Def. 6.2]{bacsar1998dynamic} defined below.

\begin{definition}\label{def:feedback-nash}
A \emph{feedback Nash equilibrium} of an \numplayers-player, general-sum, discrete-time dynamic game is an $\numplayers$-tuple of strategies~$\{\gamma^{i*}, i \in [N] \}$ with~$\control_t^{i*} = \gamma_t^{i*}(\states_t)$ that satisfies the following Nash equilibrium conditions~$\forall t \in [\horizon], i \in [\numplayers]$ for the cost in~\cref{eq:general-game-cost}:
\begin{equation}\label{eq:feedback-nash-conditions}
\begin{aligned}
    &\cost^i\bigg((\gamma^j_v)_{j = 1, v = 1}^{\numplayers, t-1}, (\gamma^{j*}_v)_{j = 1, v = t}^{\numplayers, \horizon}\bigg) \\
    &\leq \cost^i\bigg( \underbrace{(\gamma^j_v)_{j = 1, v = 1}^{\numplayers, t-1}}_{\mathrm{stages} < t}, \underbrace{\Big((\gamma^{j*}_v)_{j \neq i, v = t}^{\horizon}, \big((\gamma^i_v)_{v = t},(\gamma^{i*}_v)_{v = t+1}^{\horizon}\big)\Big)}_{\mathrm{stages} \geq t}\bigg).
\end{aligned}
\end{equation} 
\end{definition}

Intuitively, at a feedback Nash equilibrium, each player's strategy is unilaterally optimal from an arbitrary stage~$t$ onwards. 
We note that the standard feedback Nash equilibrium in~\cref{def:feedback-nash} consists of \emph{full-state} feedback strategies~$\gamma_t^{i*}$ that map \emph{all} players' states to player~$i$'s control~$\control_t^{i*} = \gamma_t^{i*}(\states_t)$.
This work seeks to find \emph{regularized} feedback Nash equilibrium strategies that approximate standard Nash equilibrium and also are sparse, \ie, depend on fewer agents' states.

\section{Approach}

\Cref{sec:lq-case} presents the proposed approach to compute regularized feedback Nash equilibria in \ac{lq} games, which is the focus of this work. \Cref{sec:non-lq-case} discusses an extension to non-\ac{lq} games.

\subsection{Regularized Dynamic Programming for Linear-Quadratic  Games}\label{sec:lq-case}

We first introduce the computation of standard feedback Nash equilibria and then discuss the modification to yield solutions with sparse inter-agent dependencies.

\subsubsection{Computation of Feedback Nash Equilibrium via \ac{dp}}\label{sec:standard-dp}

\tacinitial{
We introduce the main results of computing feedback Nash equilibria via \ac{dp}. For more details, please refer to \cite{bacsar1998dynamic,SmoothGameTheory}. 
}
A feedback Nash equilibrium solution recursively satisfies the following coupled \ac{hj} equations for value functions~$\ValueFunc_t^i(\states_t)$, $\forall i \in [\numplayers], t \in [T]$: 
\begin{equation}
\begin{aligned}    
\label{eq:hj}
    \ValueFunc_t^i(\states_t) =& \min_{\control_t^i} \ell_t^i(\states_t, \control_t) \\
    = & \min_{\control_t^i} \Big(  \StateStageCost{t} +\\
    & \ControlStageCost{t} 
    + \ValueFunc_{t+1}^i(\states_{t+1}) \Big), 
\end{aligned}
\end{equation}
where other agents play their Nash equilibrium strategies, \ie,~$\control^j_t = \gamma^{j*}(\states_t), \forall j \neq i$;~$\ValueFunc_{T+1}^i(\states_{T+1})$ is equal to the terminal cost~$\StateStageCost{T+1}$. 

\tacinitial{
The \ac{dp} procedure goes backwards in time. 
Under standard assumptions~$Q_t^i \succeq 0, R^{ii}_t \succ 0, R^{ij}_t \succeq 0, \forall i, j \in [\numplayers], i \neq j$, the objective function of each \ac{dp} iteration is strongly convex, and the feedback Nash equilibrium policy}
takes an affine form:
\begin{equation}\label{eq:linear-policy}
\control_t^{i*} = \gamma_t^{i*}(\states_t) = - P_t^{i}\states_t - \alpha_t^{i}.
\end{equation}
We refer to $P^i_t$ as a policy matrix, to which we are interested in encoding group sparsity in \cref{sec:approach-regularization}. 
The value function takes a quadratic form:
\begin{equation}\label{eq:quadratic-value-function}
    \ValueFunc_t^i(\states_t) = \half (\states_t^\top Z_t^i + 2 \eta_t^{i\top})\states_t + \beta_t^i,
\end{equation}
with~$Z_t^i, \eta_t^i, \beta_t^i$ being defined recursively:
\begin{equation}
\begin{aligned}
    \label{eq:value-func}
    Z_t^i = &Q_t^i + \sum_{j=1}^\numplayers P_t^{j\top} R_t^{ij} P_t^{j} + F_t^\top Z^i_{t+1}F_t\\
    \eta_t^i = &q_t^i + \sum_{j=1}^\numplayers (P_t^{j\top}R_t^{ij}\alpha_t^j - P_t^{j\top} r_t^{ij}) + F_t^\top (Z_{t+1}^i \omega_t +  \eta_{t+1}^i)\\
    \beta_t^i = &\half(\alpha_t^{j\top}R_t^{ij}-2r_t^{ij\top})\alpha_t^j + \half(Z_{t+1}^i\omega_t + \\
    &2\eta_{t+1}^i)^\top\omega_t + \beta_{t+1}^i,\\
\end{aligned}
\end{equation}
where:
\begin{equation*}
    \begin{aligned}
    &F_t = \At - \sum_{j=1}^\numplayers \Bjt P^j_t, \omega_t = - \sum_{j=1}^\numplayers \Bjt \alpha_t^j\\
    &Z^i_{T+1} = Q^i_{T+1}, \eta_{T+1}^i = q_{T+1}^i.
    \end{aligned}
\end{equation*}
We refer to $Z_t^i$ as a value matrix. 

\tacinitial{
At each \ac{dp} stage, a Nash equilibrium policy can be computed via solving coupled Riccati equations to compute $P_t, \alpha_t$:
}
\begin{subequations}
\label{eq:policy-linear-systems}
\begin{equation}
\label{eq:policy-linear-systems-P}
        (R_t^{ii} + B_t^{i\top}Z_{t+1}^iB_t^i ) P_t^i + B_t^{i\top}Z_{t+1}^i\sum_{j \neq i} \Bjt P_t^j = B_t^{i\top}Z_{t+1}^i\At,\\
\end{equation}
\begin{equation}
\label{eq:policy-linear-systems-alpha}
        (R_t^{ii} + B_t^{i\top}Z_{t+1}^iB_t^i)\alpha_t^i + B_t^{i\top}Z_{t+1}^i\sum_{j \neq i} \Bjt \alpha_t^j = B_t^{i\top}\eta_{t+1}^i + r_t^{ii}.
\end{equation}
\end{subequations}
We rewrite \cref{eq:policy-linear-systems-P} in a matrix form~$S_tP_t = Y_t$, which we shall use in \cref{sec:approach-regularization}:
\begin{equation}\label{eq:system-equations}
\begin{aligned}
    S_tP_t
    = \underbrace{
    \begin{bmatrix}
        B_t^{1\top}Z_{t+1}^1A_t\\
        B_t^{2\top}Z_{t+1}^2A_t\\
        \hdots\\
        B_t^{N\top}Z_{t+1}^NA_t\\
    \end{bmatrix}
    }_{Y_t}, 
\end{aligned}
\end{equation}
with~$P_t = [P_t^{1\top}, \hdots, P_t^{\numplayers\top}]^\top$ and~$S_t$ being defined as:
\begin{equation*}
\centering
\begin{aligned}
S_t = 
\begin{bmatrix}
        R_t^{11}+B_t^{1\top}Z_{t+1}^1B_t^1 &  \hdots & B_t^{1\top}Z_{t+1}^1B_t^N\\
        B_t^{2\top}Z_{t+1}^2B_t^1 & \hdots & B_t^{2\top}Z_{t+1}^2B_t^N\\
        \vdots &  \ddots & \vdots \\
        B_t^{N\top}Z_{t+1}^NB_t^1 &  \hdots & R_t^{NN} + B_t^{N\top}Z_{t+1}^NB_t^N\\
\end{bmatrix}.
\end{aligned}
\end{equation*}


\subsubsection{Regularization}\label{sec:approach-regularization}

\begin{figure*}[h]
\centering
  \includegraphics[width=0.95\linewidth]{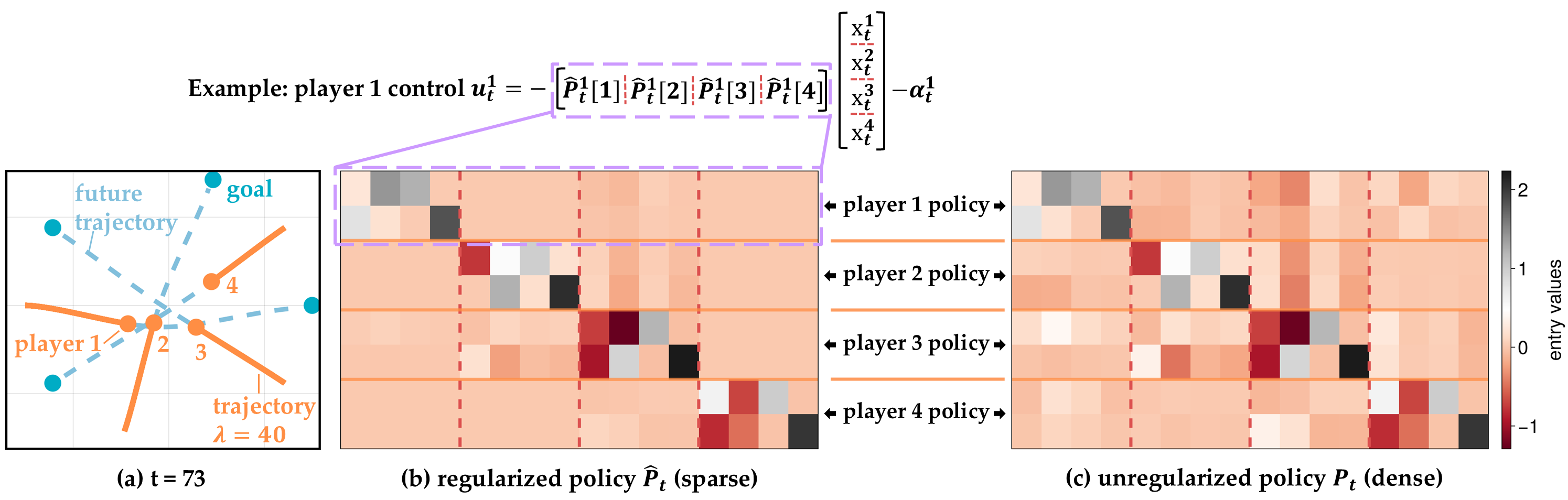}
  \caption{
 (a) Snapshot of an episode from the navigation game example.
  (b-c): Regularized and standard Nash equilibrium policy matrices, denoted as $\Pht$ in \cref{eq:reg-problem} and $P_t$ in \cref{eq:system-equations}.  
  A policy matrix~$P_t = [P_t^{1\top}, \hdots, P_t^{\numplayers\top}]^\top$, aggregated over players, maps all players' states to joint controls: $\control_t = - P_t \states_t - \alpha_t$. 
  Each policy matrix is divided into $\numplayers \times \numplayers$ blocks.
  }
  \label{fig:navigation-game-1}
\end{figure*}

At each \ac{dp} stage, solving the system in \cref{eq:system-equations} exactly computes the feedback parts of the \emph{standard} Nash equilibrium strategies. 
To compute sparse policies~$\Pht$, we propose to solve a \emph{regularized} problem:
\begin{equation}\label{eq:reg-problem}
    \min_{\Pht} \frac{1}{2} \| \St \Pht - \Yt \|_F^2 + \sum_{i,j} \lambda_{i,j} \|\Pht^i[j]\|_F, 
\end{equation}
where $\normf{\cdot}$ denotes matrix Frobenius norm and $\Pht^i[j]$ denotes a \emph{block} in the policy matrix that maps the \jth player's states to the \ith player's controls. 
For example, in a~4-player game, a policy matrix $P_t$ or $\Pht$ shown via a heatmap in \cref{fig:navigation-game-1} is divided into $4 \times 4$ blocks. 
Throughout the manuscript, we use a hat notation to denote quantities computed in the recursion with policy sparsity regularization, \eg, $\Pht, \hat{Z}_t$. 
$\lambda_{i,j}$ denotes a weighting constant that determines the regularization strength for block~$(i,j)$. 
Setting $\lambda_{i,j} = 0, \forall i,j \in [\numplayers]$ recovers a standard feedback Nash equilibrium solution. 
For a block~$(i,j)$, as the user increases~$\lambda_{i,j}$, the values in the block are penalized more until the entire block is \say{zeroed out}. 
We note that we choose~$\lambda_{i,i} = 0, \forall i \in [\numplayers]$ and~$\lambda_{i,j} = \lambda, \forall i \neq j, \lambda \in \R_{\geq 0}$, to not discourage the players' strategies from depending on their own states and to penalize other blocks evenly. 
Hence, by solving the problem in~\cref{eq:reg-problem}, the computed policy can automatically choose to depend on fewer agents' states. 
Solving the problem in~\cref{eq:reg-problem} can be interpreted as solving the problem in~\cref{eq:hj} with sparsity regularization.

The problem in~\cref{eq:reg-problem} is a group Lasso problem, which was initially proposed in~\cite{yuan2006model} and has been extended and widely applied to grouped variable selection, \eg, in high-dimensional statistics~\cite{buhlmann2011statistics} and signal processing~\cite{lv2011group}. More specifically in our case, since~$\lambda_{i,i} = 0, \forall i \in [\numplayers]$ in~\cref{eq:reg-problem}, our problem is termed an adaptive group Lasso problem \cite[Ch.4]{buhlmann2011statistics}. 

Importantly, the problem in~\cref{eq:reg-problem} encourages sparsity in the solution at a \emph{group} level. 
The entries in a group all remain non-zero or get zeroed out together. 
The problem is \emph{convex}, and can be solved using established algorithms for group Lasso, \eg, via block coordinate descent~\cite{yuan2006model,buhlmann2011statistics,meier2008group} or a projected gradient method~\cite{kim2006blockwise}. 
We also show in Appendix that this problem can be cast as a conic program and effectively solved via off-the-shelf conic optimization solvers.

Hence, at each dynamic programming iteration~$t$ described in~\cref{sec:standard-dp}, instead of solving the system in~\cref{eq:system-equations} exactly, we solve the problem in~\cref{eq:reg-problem} to compute regularized Nash equilibrium strategies. 
Using the regularized strategies $\hat{\gamma}^{i*}_t$, we then compute value functions~$\hat{\ValueFunc}_t^i(\states_t)$ that correspond to state values for the \emph{sparse} policies. 
We repeat this process backwards for every stage~$t \in [T]$.

\begin{remark}
Note that our approach in~\cref{eq:reg-problem} regularizes a specific step in the \ac{dp} associated with \ac{lq} feedback games and maintains convexity at each step. 
In contrast, prior efforts \cite{lin2013design,dorfler2014sparsity,wytock2013fast,park2020structured} (in the single-agent context) regularize the \emph{entire} policy optimization objective and solve a nonconvex problem. 
\tacinitial{
In general, extending direct policy optimization methods to general-sum games can face convergence challenges~\cite{mazumdar2020policy}.}
\end{remark}

\subsubsection{Convergence of the Regularized Dynamic Program}\label{sec:theory}

In this section, we analyze the convergence of the proposed \ac{dp} scheme on the regularized policy matrix $\Pht$ and value matrix $\hat{Z}_t$ to a neighborhood of the Nash equilibrium. 

We denote an unregularized solution to the system in~\cref{eq:system-equations} as~$P_t$ and a regularized solution to the problem in~\cref{eq:reg-problem} as~$\Pht$. 
We define the difference between the two solutions as $\Delta P_t := \Pht - P_t$. 
We also denote the minimum singular value of a matrix as $\sigma_{\mathrm{min}(\cdot)}$. 
When we consider the asymptotic behavior of finite-horizon games, we shall fix the horizon $\horizon$ as a finite constant and allow $t \downarrow -\infty$.

We refer to the recursion on the value matrix $Z_t^i$ in \cref{eq:value-func} as  \say{unregularized coupled Riccati recursion}. 
The proposed approach instead solves a regularized problem in \cref{eq:reg-problem} at each iteration and therefore yields a \say{regularized coupled Riccati recursion}:
\begin{equation}\label{eq:regularized-riccati}
    \hat{Z}_t^i = Q^i + \sum_{j=1}^\numplayers \Pht^{j\top} R^{ij} \Pht^{j} + \Fht^\top \hat{Z}^i_{t+1}\Fht, 
\end{equation}
where $\Fht = F_t - \sum_{j = 1 }^\numplayers B^j\Delta P^j_t$. 
This section analyzes the difference between the regularized and unregularized Riccati recursions. 
Note that the analysis will also use the notion of \say{infinite-horizon coupled Riccati equation}: 
\begin{equation}\label{eq:inf-horizon-riccati}
    Z^{i*} = Q^i + \sum_{j=1}^\numplayers P^{j*\top} R^{ij} P^{j*} + F^{*\top} Z^{i*}F^*,
\end{equation}
which is defined similarly as \cref{eq:value-func} but for settings when the horizon of the game in \cref{def:lq-game} is infinite. 
With a slight abuse of notation, we note that here the absence of subscript denotes that the value of $Z^{i*}$ is static across time. 
In the analysis in this section, we focus on the settings where the problem data, i.e. the collection $(Q^i_t, q^i_t, R^{ij}_t, r^{ij}_t)_{i,j \in [\numplayers], t \leq \horizon+1}$, for the game in \cref{def:lq-game} are time-invariant, which is standard when analyzing convergence of Riccati recursions \cite[Ch.6]{bacsar1998dynamic} and holds true for all of our experimental examples in~\cref{sec:results}.

We first analyze and bound the difference between solving the regularized problem in~\cref{eq:reg-problem} compared to solving the system in~\cref{eq:system-equations} exactly for a single stage. 
However, tracking the differences propagated through the Riccati recursion in~\cref{eq:regularized-riccati} is generally challenging. 
We therefore take a dynamical system perspective and prove that under certain mild conditions, the regularized Riccati recursion converges to a neighborhood of the unregularized Riccati recursion, \ie, the regularization does not cause a diverging error. 
Under a mild assumption, the single-step $\Delta P_t$ can be explicitly upper bounded. 

\begin{assumption}\label{assumption:unique-nash}
    We assume that the unregularized game in~\cref{def:lq-game} has a unique Nash equilibrium, \ie, $S_t$ in~\cref{eq:system-equations} is invertible.
\end{assumption}

This assumption is mild and is a common setup in the literature \cite[Ch.6]{bacsar1998dynamic}. 
In our experiments in \cref{sec:results}, we do not have an issue satisfying this invertibility requirement of $S_t$. 

\begin{lemma}\label{lemma:one-step-bound}
    Let \cref{assumption:unique-nash} hold and let $P_t$ solve \cref{eq:system-equations} and $\Pht$ solve the problem in \cref{eq:reg-problem} with the same problem data $S_t, Y_t$. 
    Then, the difference $\Delta P_t$ between $\Pht$ and $P_t$ can be upper bounded as:
    \begin{equation}\label{eq:upper-bound}
        \normf{\Delta P_t} \leq \frac{\sum_{i,j \in [\numplayers]}\lambda_{i,j}}{\sigma^2_{\mathrm{min}}(S_t)}.
    \end{equation} 
\end{lemma}

By \cref{assumption:unique-nash}, the original game in~\cref{def:lq-game} has a unique solution, $S_t$ in \cref{eq:system-equations} is of full rank and has positive singular values. Hence, the upper bound in \cref{eq:upper-bound} is finite when $\lambda_{i,j} < \infty$ for all~$i,j \in [\numplayers]$.
We prove~\cref{lemma:one-step-bound} by casting the problem of solving~\cref{eq:system-equations} as a least-squares optimization problem and compare the optimality conditions of the unregularized problem and the nonsmooth regularized problem. 
We give the complete proof in Appendix. 

\Cref{lemma:one-step-bound} provides a bound on~$\Delta P_t$ that shrinks to zero as~$\lambda$ decreases. 
While \cref{lemma:one-step-bound} accounts for the situation where the problem data~$S_t,Y_t$ are the same for the unregularized and regularized problems, this is not the case when the difference caused by regularization propagates through the coupled Riccati recursion in \cref{eq:regularized-riccati}. 
That is, the regularized \ac{dp} recursion will ultimately have problem data~$\hat{S}_t, \hat{Y}_t$ that are different from the unregularized system in \cref{eq:system-equations}. 
Under another mild assumption, we can overcome this challenge by leveraging dynamical system theory. 
\begin{assumption}\label{assumption:convergence-riccati}
We assume that the steady-state matrix $Z^*$ in \cref{eq:inf-horizon-riccati} is a \ac{las} fixed point of the finite-horizon Riccati recursion for~$Z_t$ in \cref{eq:value-func}. 
The absence of the superscripts denotes aggregation over all the players. 

\end{assumption}

By \cref{assumption:convergence-riccati}, $Z_t$ in \cref{eq:value-func} locally converges to $Z^*$ in \cref{eq:inf-horizon-riccati} at the limit $t \rightarrow -\infty$. 
This requirement is not restrictive,
as taking the limit of the finite-horizon Riccati sequence is a common approach to finding Nash equilibria for infinite-horizon \ac{lq} games~\cite[Ch.6]{bacsar1998dynamic}. 
Specific conditions to guarantee that the infinite-horizon Riccati solution is a \ac{las} fixed point can be obtained for \ac{lqr} and zero-sum \ac{lq} games, but it is challenging to get such conditions for general-sum \ac{lq} games~\cite[Ch.6]{bacsar1998dynamic}. 

We define the difference between \cref{eq:value-func} and \cref{eq:inf-horizon-riccati} as a dynamical system~$\StateRiccatiAll_t := Z_t - Z^* = \DynRiccati(\StateRiccatiAll_{t+1}, 0)$ with zero input and the origin as a \ac{las} point. 
Regularization using the proposed scheme in~\cref{eq:reg-problem} injects a disturbance~$\delta_t = \Delta P_t$ to the system at each stage and yields a perturbed system~$\StateRiccatiAllReg_t = \DynRiccati(\StateRiccatiAllReg_{t+1}, \delta_t)$, which corresponds to the difference between \cref{eq:regularized-riccati} and \cref{eq:inf-horizon-riccati}. 
To this end, we provide the following theorem and corollary. 

\begin{theorem}\label{theorem:iss}
Let \cref{assumption:unique-nash} and \cref{assumption:convergence-riccati} hold. Then, the dynamical system~$\StateRiccatiAllReg_t = \DynRiccati(\StateRiccatiAllReg_{t+1}, \delta_t)$, with the difference $\StateRiccatiAllReg_t := \hat{Z}_t - Z^*$ being the state and the policy deviation $\Delta P_t$ caused by regularization in~\cref{eq:reg-problem} 
as disturbances $\delta_t$, is locally \ac{iss}.
That is, there exist a class~$\mathcal{KL}$ function~$\RiccatiTransient$ and a class $\mathcal{K}$ function $\BallRiccati$ such that: 
\begin{equation}
    \|\StateRiccatiAllReg_t\|_F \leq \RiccatiTransient\Big(\|\StateRiccatiAllReg_{\horizon + 1}\|_F, \horizon - t\Big) + \BallRiccati(\norm{\delta_t}_{\infty}), \forall t \in (-\infty, \horizon],
\end{equation}
for disturbances such that~$\sup_{t \leq\tau\leq \horizon}\normf{\delta_\tau} = \norm{\delta_t}_{\infty} < \delta, \delta \in \R_{>0}$ and $\|\StateRiccatiAllReg_{\horizon + 1}\|_F \leq z, z \in \R_{>0}$.
\end{theorem}

\begin{corollary}\label{corollary:iss}
Let \cref{assumption:unique-nash} and \cref{assumption:convergence-riccati} hold and $\|\StateRiccatiAllReg_{\horizon + 1}\|_F \leq z, z \in \R_{>0}$. 
Then, for all $\epsilon > 0$, there exists $\xi > 0$, such that for all $\lambda_{i,j}$ satisfying $\sum_{i,j \in [\numplayers]}\lambda_{i,j} \leq \xi$, the regularized coupled Riccati recursion approaches the
ball with radius $\epsilon$:
\begin{equation}\label{eq:ultimate-bound}
    \limsup_{t \rightarrow -\infty} \normf{\hat{Z}_t - Z^{*}} \leq \epsilon.
\end{equation}
\end{corollary}

\begin{remark}
    Given the convergence of $\hat{Z}_t^i$, the convergence of $\Pht$ also follows. 
    Particularly, from \cref{eq:system-equations}, we notice that $S_t$ and $Y_t$ both depend linearly on $Z_{t+1}^i$ and \cref{eq:system-equations} is a linear system of equations with respect to $P_t$. Therefore, one can relate the ultimate bound of $\norm{\hat{P}_t - P^*}_F$ back to the ultimate bound of $\norm{\StateRiccatiAllReg_t}_F$ in \cref{eq:ultimate-bound} using linear system perturbation theory~\cite[Ch.6]{horn2012matrix} and the bound in \cref{lemma:one-step-bound}. 
\end{remark}

We provide a complete proof of \cref{theorem:iss,corollary:iss} in the Appendix. 
\cref{theorem:iss} and \cref{corollary:iss} suggest that the regularized Riccati recursion will converge to a neighborhood of the infinite-horizon Riccati equation under bounded disturbances 
caused by the regularization. 
Our analysis shows the asymptotic tradeoff between players’ costs and policy sparsity, and ensures that the difference in cost-to-go induced by the group-sparsity regularization of \cref{eq:reg-problem} is bounded under bounded regularization weights $\sum_{i,j \in [\numplayers]}\lambda_{i,j} \leq \xi$. 
For large regularization such that $\sum_{i,j \in [\numplayers]}\lambda_{i,j} > \xi$, we also empirically observe convergence of the Riccati recursion in \cref{sec:formation-game}. 
Our conjecture is that the sparsity regularization added to off-diagonal blocks in policy matrices causes the regularized Riccati recursion to be more \say{decoupled} between players and the dynamical system~$\StateRiccatiAllReg_t = \DynRiccati(\StateRiccatiAllReg_{t+1}, \delta_t)$ to be more stable. 
However, a more detailed analysis is out of the scope of this work. 
We also empirically show the convergent behaviors for the unregularized and regularized Riccati recursions in~\cref{sec:formation-game}.

\subsection{Non-\ac{lq} Dynamic Games}\label{sec:non-lq-case}

The proposed approach focuses on \ac{lq} dynamic games. 
Nonetheless, our \ac{dp}-based scheme makes it possible to extend to non-\ac{lq} games, \ie, settings where players can have nonquadratic costs and nonlinear dynamics. 
We employ an iterative algorithm~\cite{fridovich2020efficient} shown in~\cref{algo:ilqgames} that repeatedly finds \ac{lq} approximations of the original dynamic game and solves the approximating \ac{lq} games with regularization in~\cref{eq:reg-problem} to obtain sparse, approximate feedback Nash equilibrium strategies. 
\tacinitial{
This type of iterative \ac{lq} algorithm has been studied in the context of standard (unregularized) feedback Nash games \cite{fridovich2020efficient,laine2021computation,di2019newton}, and in \cite{laine2021computation} it is shown that fixed points of these algorithms approximately satisfy the first-order optimality conditions for all players which correspond to the feedback Nash equilibrium concept in \cref{def:feedback-nash}. 
We do not provide convergence results for non-\ac{lq} games. 
For a more detailed analysis of convergence and open challenges, please refer to \cite{laine2021computation}. 
}

\begin{algorithm}[h]
\caption{Iterative \ac{lq} Games with Group Sparsity}
\label{algo:ilqgames}
\SetAlgoLined
{%
\DontPrintSemicolon
\textbf{Input:} initial state~$\states_1$, game dynamics~$\{f_t\}_{t \in [\horizon]}$, game costs~$\{J^i\}_{i \in [\numplayers]}$\\
\textbf{Initialization:} Initial control strategies~$\{\hat{\gamma}^{i}\}_{i \in [\numplayers]}$\\
\tcc{Iterative \ac{lq} Approximations}
\While{$\mathrm{not~converged}$}{
    $(\states_{1:\horizon+1}, \control_{1:\horizon}) \gets$ \\
    $\qquad \operatorname{ForwardSimulate}(\{\hat{\gamma}^{i}\}_{i \in [\numplayers]}, \states_1, \{f_t\}_{t \in [\horizon]})$\\
    $(A_t, B_t)_{t \in [\horizon]} \gets$ \\
    $\qquad \operatorname{LinearApproximation}(\states_{1:\horizon+1}, \control_{1:\horizon}, \{f_t\}_{t \in [\horizon]})$\\
    $(Q^i_t, q^i_t, R^{ij}_t, r^{ij}_t)_{i,j \in [\numplayers], t \in [\horizon+1]} \gets$ \\
    $\qquad \operatorname{QuadraticApproximation}(\states_{1:\horizon+1}, \control_{1:\horizon}, \{J^i\}_{i \in [\numplayers]})$\\
    $\{\Bar{\hat{\gamma}}^{i}\}_{i \in [\numplayers]} \gets$\\ $\qquad \operatorname{SolveRegularizedLQGame}(A_t, B_t, Q^i_t, q^i_t, R^{ij}_t, r^{ij}_t)$
    \tcp{regularization, eq.~(\ref{eq:reg-problem})}
    $\{\hat{\gamma}^{i}\}_{i \in [\numplayers]} \gets \operatorname{UpdateStrategies}(\{\hat{\gamma}^{i}\}_{i \in [\numplayers]}, \{\Bar{\hat{\gamma}}^{i}\}_{i \in [\numplayers]})$
}
\textbf{return} $\{\hat{\gamma}^{i}\}_{i \in [\numplayers]}$
}
\end{algorithm}

\section{Experimental Results}\label{sec:results}

We extensively evaluate the proposed approach both in non-\ac{lq} and \ac{lq} dynamic game scenarios to support our key claims made in~\cref{sec:intro}.
\footnote{Supplementary video can be found on our project website.} 

\subsection{Multi-Agent Navigation Game}
\label{sec:multi-agent-navigation}

First, we test the proposed approach in a multi-agent navigation game. 
This is a non-\ac{lq} game and provides an intuitive setting to demonstrate the computed sparsity pattern by our approach over time. We will analyze the performance of the proposed approach more closely in an \ac{lq} setting in~\cref{sec:formation-game}.

\subsubsection{Experiment Setup and Claims}

As is shown in~\cref{fig:navigation-game-1} (a), four agents start from the initial positions and drive to their individual goals. 
All the agents wish to go to their goals as efficiently as possible without too much control efforts while avoiding collision with one another. 
Hence, they need to compete and find underlying Nash equilibrium strategies. Each agent is modelled using unicycle dynamics with states being 2D position, orientation, and velocity~$x_t^i = [p_{x,t}^i, p_{y,t}^i, \phi_t^i, v_t^i]^\top$ and controls being angular and longitudinal acceleration~$u_t^i = [\omega_t^i, a_t^i]^\top$. 
We provide the definition of players' running costs in Appendix.

This experiment is designed to support the following claims:
\begin{itemize}
    \item \textbf{C1.} The proposed approach identifies influential agents and yields more sparse feedback policies than standard Nash equilibrium solutions. 
    \item \textbf{C2.} The algorithm gives different levels of sparsity as the user varies the
    regularization strength.
    \item \textbf{C3.} Via an iterative scheme~\cite{fridovich2020efficient}, the proposed approach can be extended to non-\ac{lq} dynamic games.
\end{itemize}

\subsubsection{Results}

\noindent\textbf{Sparsity Pattern.} 
\Cref{fig:navigation-game-1} 
shows a snapshot of the navigation game and the sparsity pattern in the policy obtained using \cref{algo:ilqgames}. 
On a high level, as is shown in the heatmaps, the proposed approach computes a more sparse feedback policy in \cref{fig:navigation-game-1}b than the standard Nash equilibrium in \cref{fig:navigation-game-1}c. 
In this four-player game, a policy matrix~$P_t$ has~16 blocks corresponding to feedback maps from all players' states to all players' controls.

\begin{figure}
    \begin{subfigure}{0.49\linewidth}
        \includegraphics[width=1\linewidth]{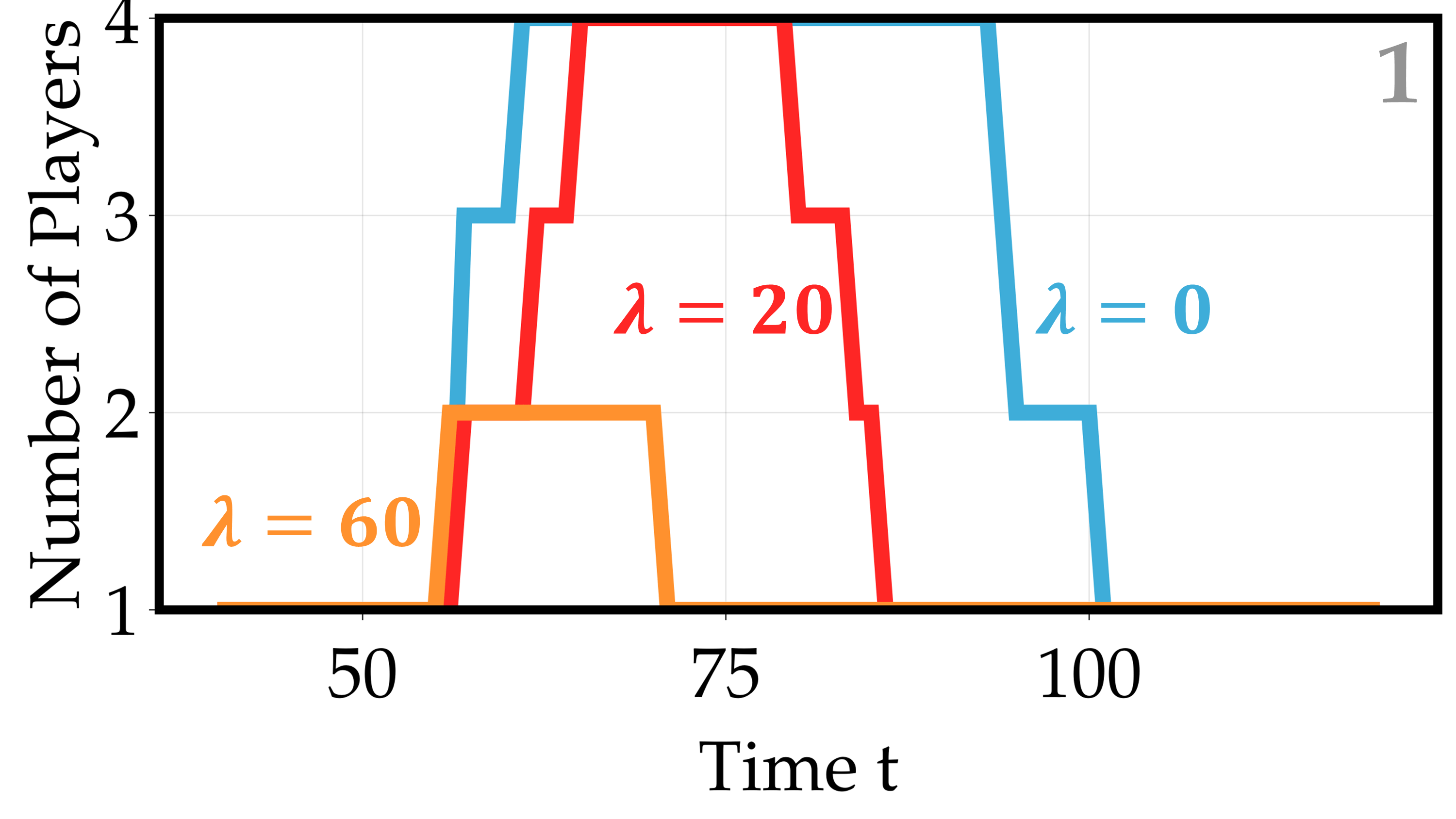}
    \end{subfigure}
    \begin{subfigure}{0.49\linewidth}
        \includegraphics[width=1\linewidth]{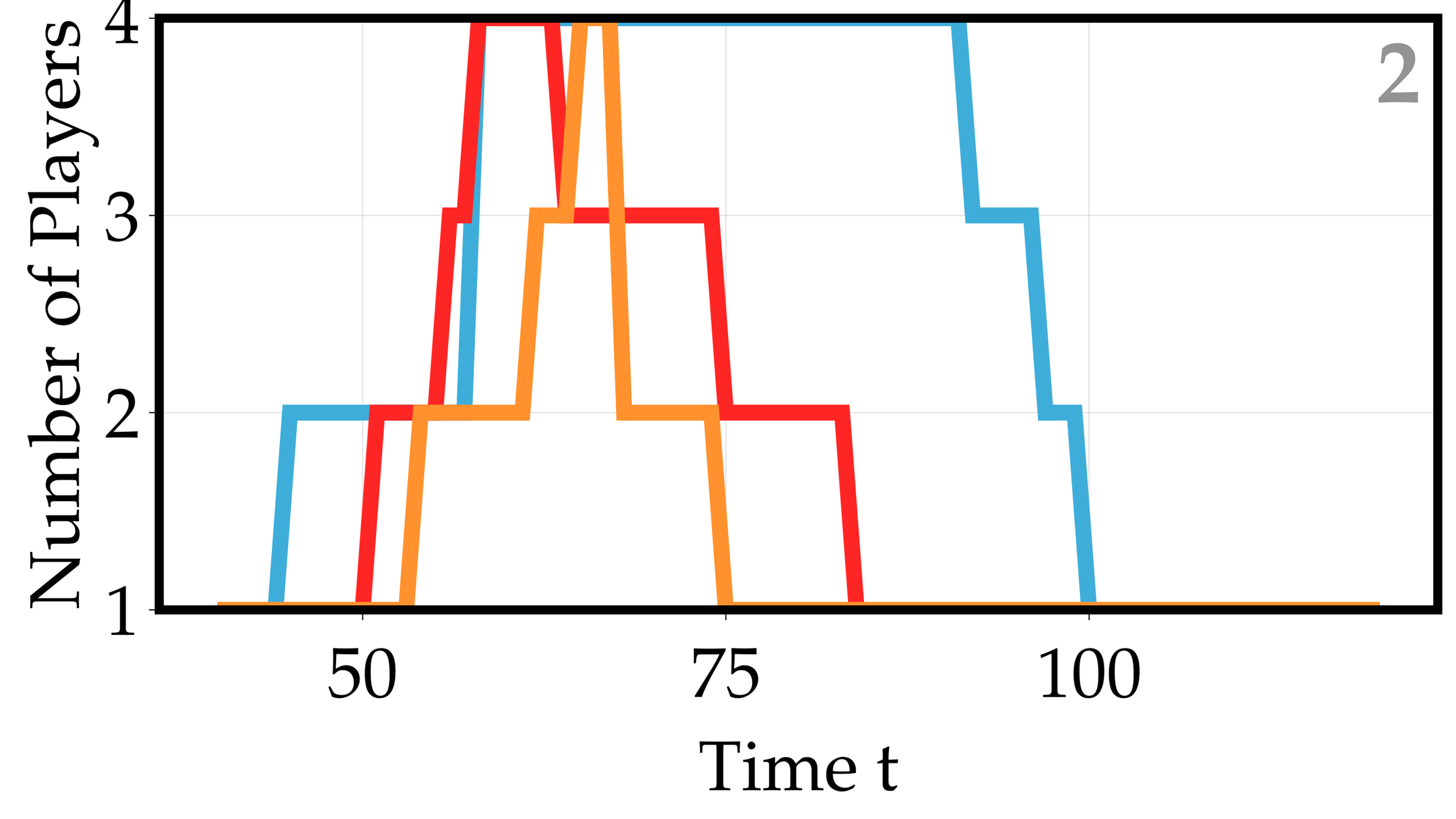}
    \end{subfigure}
    \begin{subfigure}{0.49\linewidth}
        \includegraphics[width=1\linewidth]{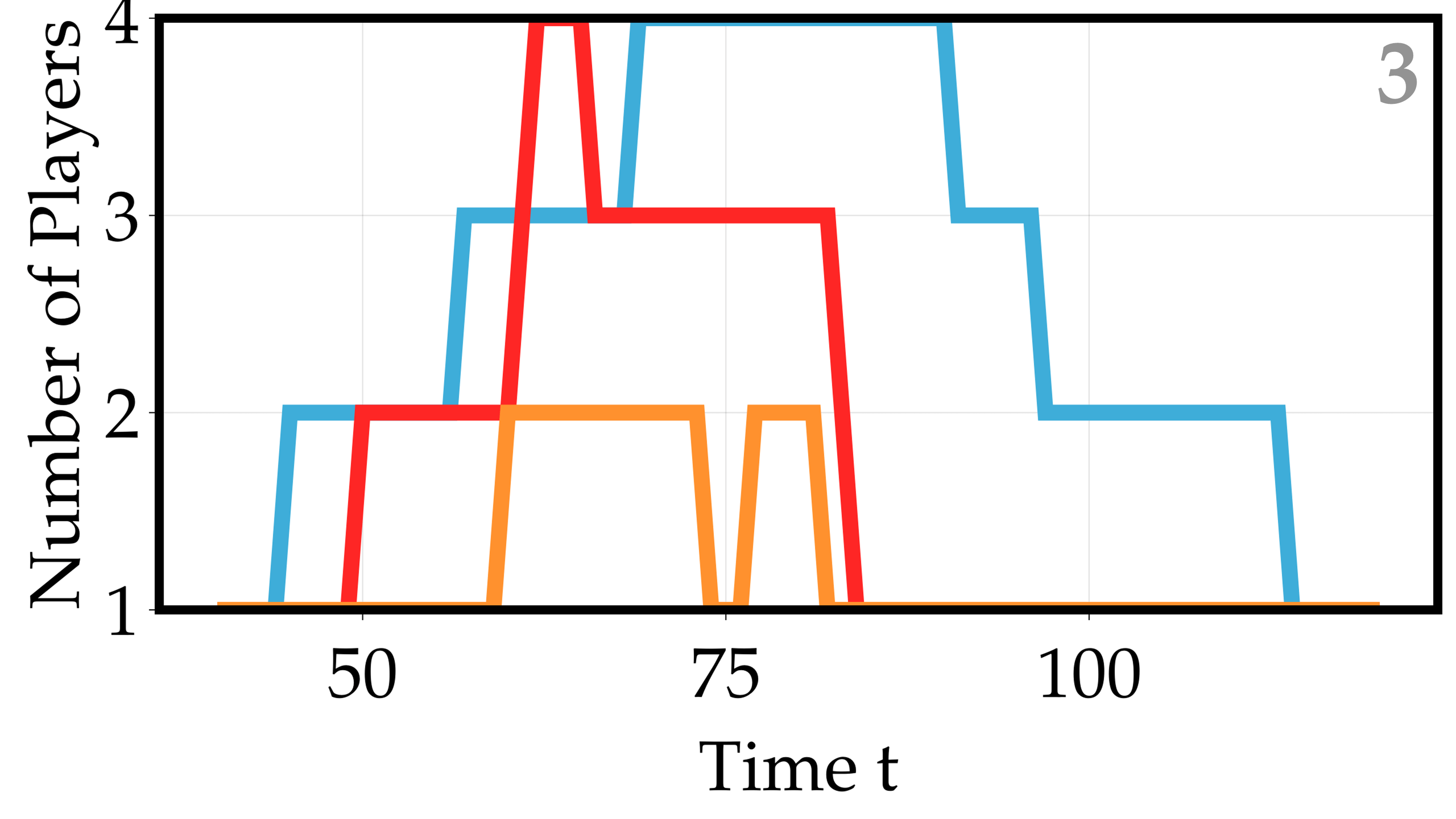}
    \end{subfigure}
    \begin{subfigure}{0.49\linewidth}
        \includegraphics[width=1\linewidth]{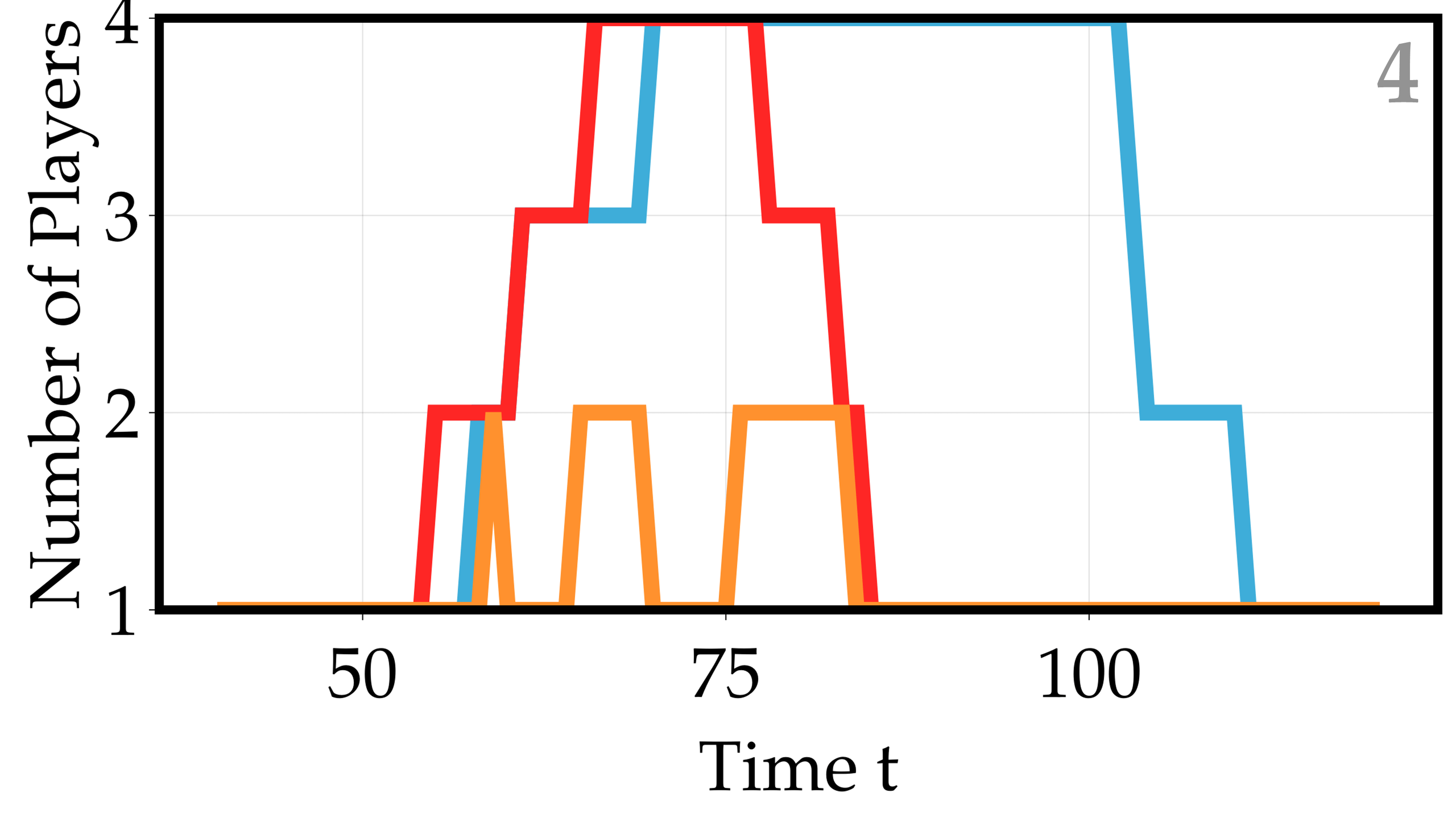}
    \end{subfigure}
    \caption{The number of nonzero blocks over time contained in policies of players 1-4 with different regularization levels.
    }
  \label{fig:navigation-sparsity}
\end{figure}

\begin{figure*}[h]
\centering
  \includegraphics[width=0.85\linewidth]{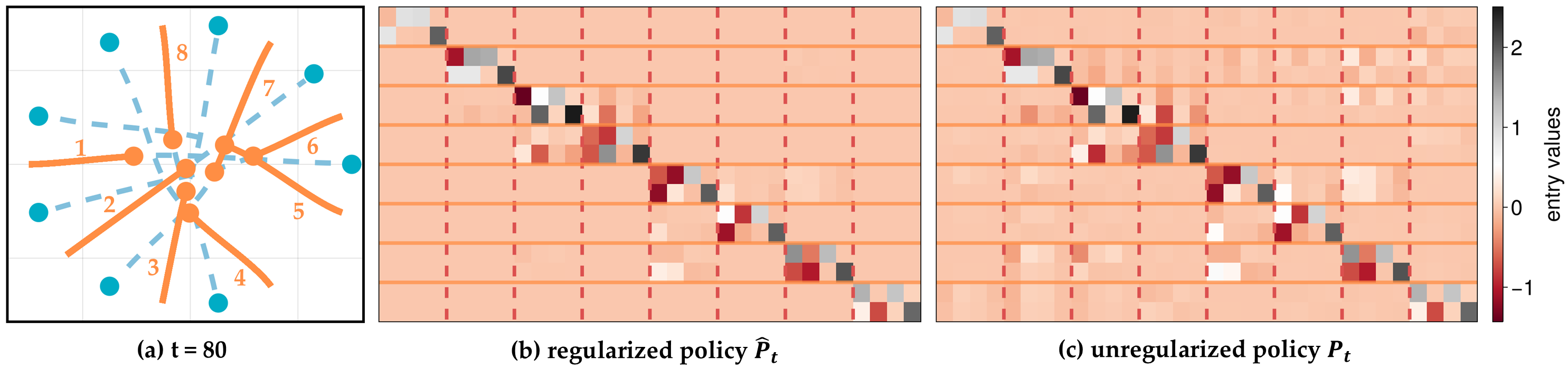}
  \caption{Snapshot of the proposed approach for an 8-player navigation game and corresponding policy matrices. 
  }
  \label{fig:navigation-sparsity-8players}
\end{figure*}

At time step~$t=73$, because agent~$2$ has already passed by agent~$1$ and their interaction has been resolved, the regularized policy by our approach chooses to zero out the blocks~$(1,2)$ and~$(2,1)$ from the policy matrix. 
Similarly, since agent~$4$ is relatively far from agents~$1$ and~$2$, the regularized policy zeros out the corresponding blocks as well.

\noindent\textbf{Different Regularization Levels.} \Cref{fig:navigation-sparsity} compares the \say{density} of the resulting feedback policies from three levels of regularization strength for the same four-player navigation game in~\cref{fig:navigation-game-1}. 
The unregularized Nash equilibrium policy matrix has the highest density in general, as the policy requires state information from the most agents over time. 
As expected, the proposed approach gives generally more sparse feedback policies with higher regularization strength. 
Note that this increased sparsity does not appear strictly all the time in~\cref{fig:navigation-sparsity}. 
This is because the evaluated non-\ac{lq} game is known to have multiple local Nash equilibrium solutions~\cite{peters2020inference}. 
In the proposed \ac{dp} approach, each stage's sparsity depends on sparsity in later stages. 
Based on the selected regularization level, the proposed approach finds different local solutions where the players pass by one another in different orders.%

\noindent\textbf{Scalability.} \Cref{fig:navigation-sparsity-8players} demonstrates our approach for the navigation game with more players, for which encoding sparsity into the computed solution is typically an even more desirable feature.

We note that all the episodes above with different regularization levels result in a collision-free interaction among the agents with only a modest change in players' costs. We defer a more detailed performance study of our regularization approach to an \ac{lq} example in~\cref{sec:formation-game}, which has a unique Nash equilibrium solution.

\emph{Hence, the results above support the claims \textbf{C1-3}.}

\subsection{Multi-Robot Formation Game}
\label{sec:formation-game}

This section evaluates the proposed approach in an \ac{lq} formation game to provide a more detailed performance analysis. 

\begin{figure}[h]
    \centering
    \begin{subfigure}[b]{0.45\linewidth}  %
        \centering
        \includegraphics[width=\linewidth]{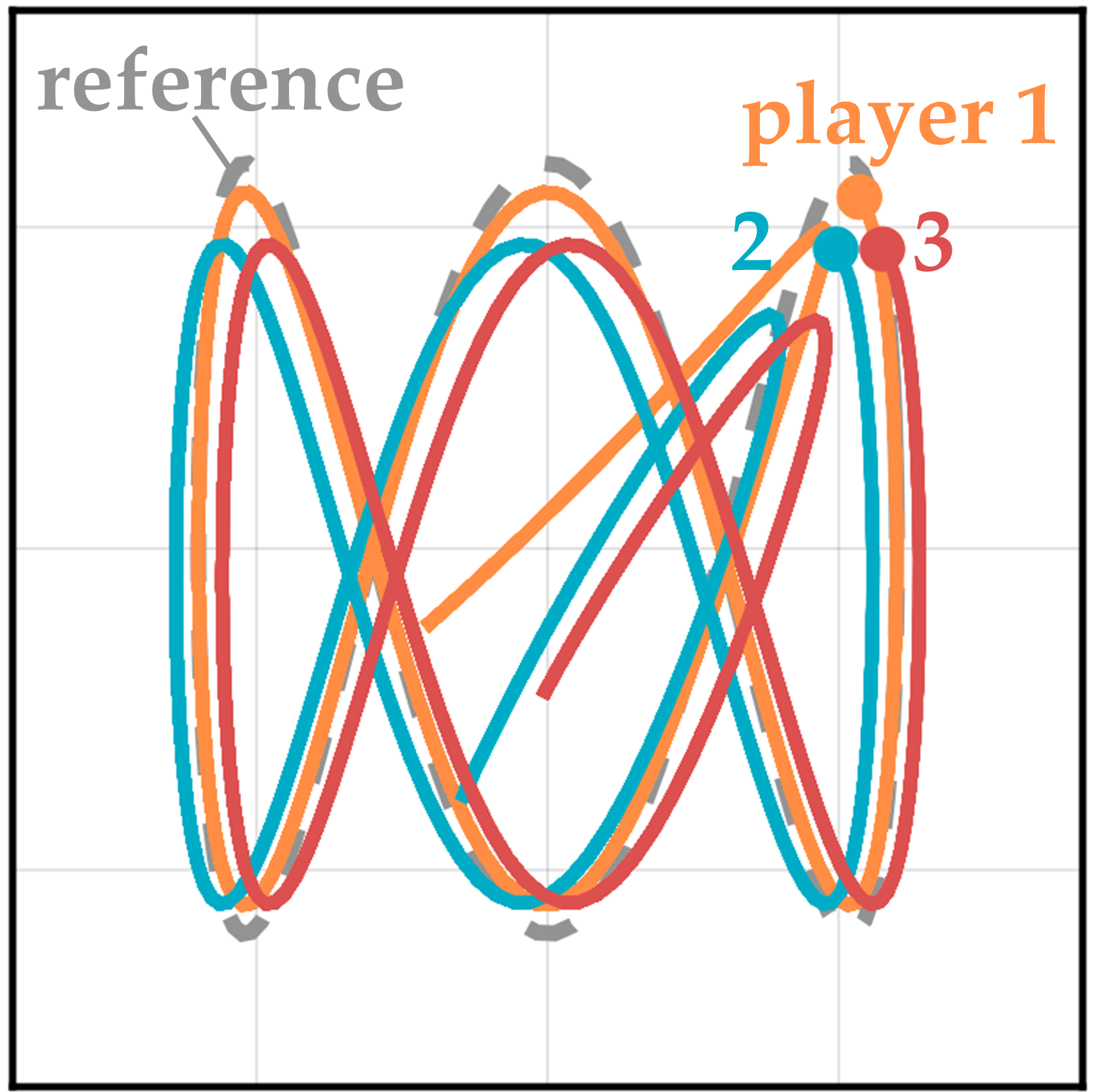}
        \caption{}  %
        \label{fig:formation-game}
    \end{subfigure}%
    \hspace{0.05\linewidth}  %
    \begin{subfigure}[b]{0.45\linewidth}  %
        \centering
        \includegraphics[width=\linewidth]{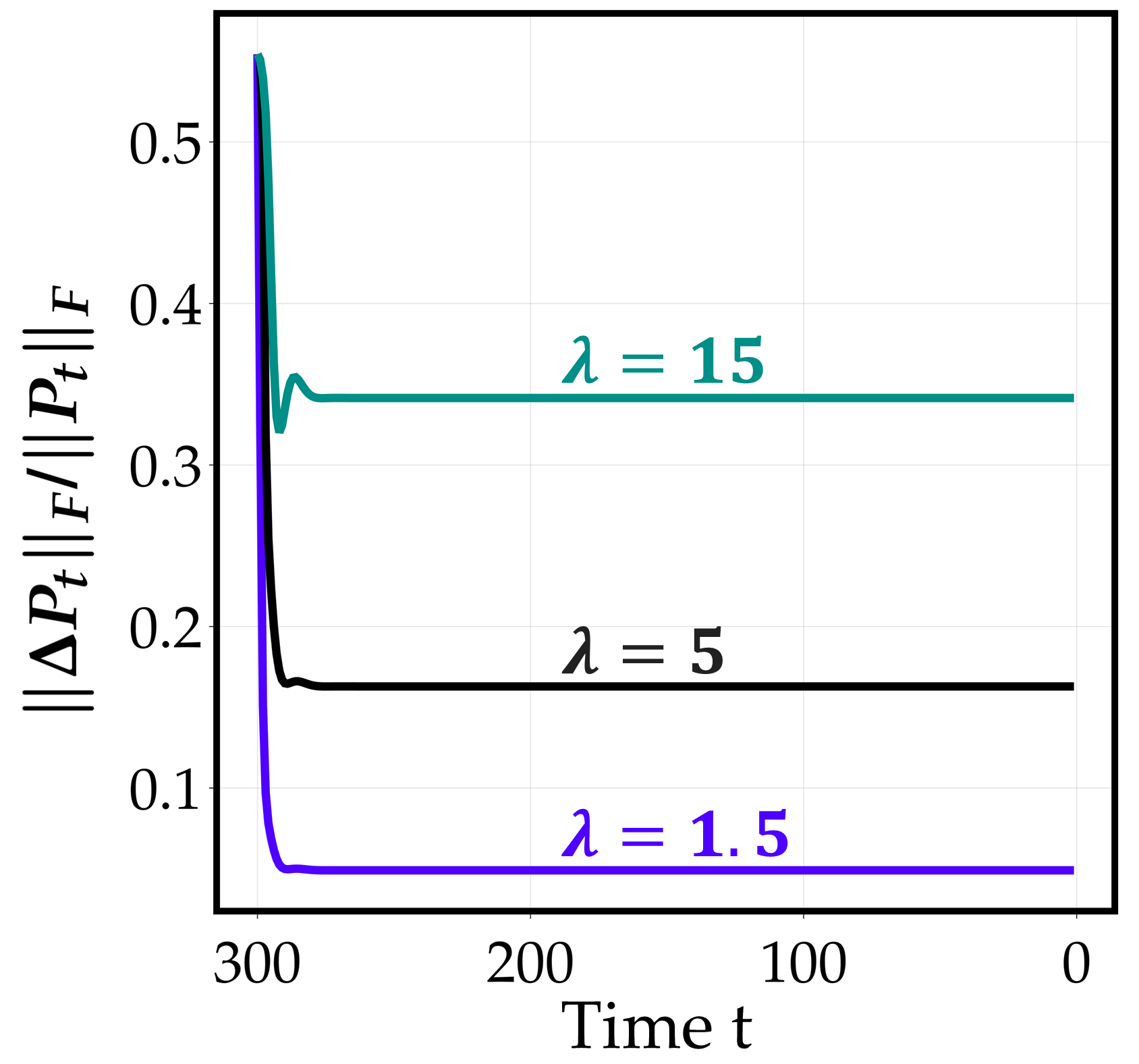}
        \caption{}  %
        \label{fig:dp-recursion}
    \end{subfigure}
    \caption{(a) A three-player formation game: player 1 tracks a reference trajectory shown in grey while players 2 and 3 maintain a formation with respect to player 1.
    (b) Convergence of regularized policies with different $\lambda$.}
\end{figure}

\subsubsection{Experiment Setup and Claims}

\Cref{fig:formation-game} shows the following scenario: player~1 tracks a predefined sinusoidal trajectory, while player~2 is tasked to maintain a relative position with players~1 and~3, and likewise for player~3. Hence, player~1's cost is independent of the other players, while costs for players 2 and 3 are defined by other players' positions. As a result, the players need to maintain a formation together while tracking the sinusoidal trajectory. Each player's dynamics are modelled as a planar double integrator with states being position and velocity~$x_t^i = [p_{x,t}^i, v_{x,t}^i, p_{y,t}^i, v_{y,t}^i]^\top$ and controls being acceleration~$u_t^i = [a_{x,t}^i, a_{y,t}^i]$. We provide the definition of players' costs in Appendix.

This experiment is designed to support the following claim:

\begin{itemize}
    \item \textbf{C4.} For all interacting agents whose costs are coupled with other agents’ states, the sparse strategies computed by our approach improve robustness and task performance over standard Nash equilibrium solutions when agents only have inaccurate (\eg, noisy) estimates of other agents’ states. 
\end{itemize}

\begin{figure*}[h]
\centering
  \includegraphics[width=0.95\linewidth]{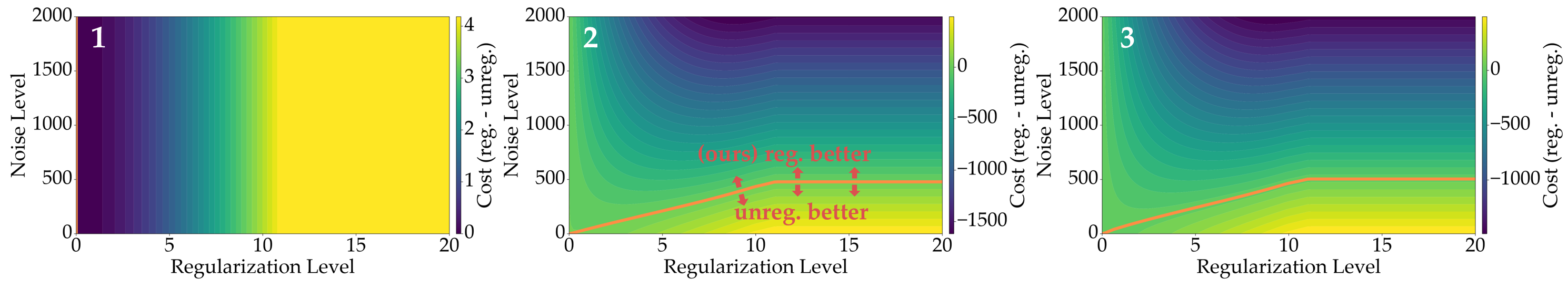}
  \caption{Costs of players 1-3 subtracted by standard Nash equilibrium costs in a formation game, averaged over 100 random initial players' positions, for varying noise and regularization levels. In subplots 2-3, the orange lines indicate the zero level set. 
  }
  \label{fig:formation-game-cost}
\end{figure*}

\begin{figure*}[h]
\centering
  \includegraphics[width=0.8\linewidth]{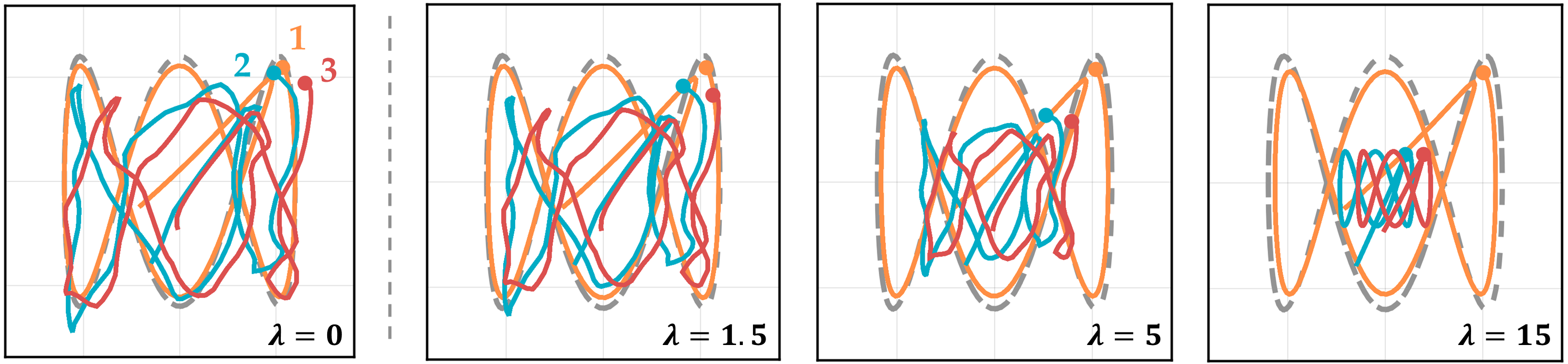}
  \caption{Formation game trajectories under noise level~520 with different regularization levels; left to right: $\lambda = 0, 1.5, 5  ~(\text{optimal}), 15$.}
  \label{fig:formation-game-qualitative}
\end{figure*}

\subsubsection{Results}

\Cref{fig:formation-game} illustrates a resulting trajectory generated by a standard Nash equilibrium strategy, serving as the ground truth behavior.
The three players follow the task trajectory from their initial positions, forming a triangular formation along the path.
\Cref{fig:dp-recursion} shows the convergence of the policy deviation, $\Delta P_t$, across \ac{dp} iterations for various $\lambda$ values. 
Notably, the regularized \ac{dp} recursion converges even for a high $\lambda = 15$, which completely zeros out the off-diagonal blocks of the policy matrices. 
\emph{This convergence validates the theoretical results of \cref{theorem:iss,corollary:iss}.}

\noindent\textbf{Monte Carlo Study.} To better understand our regularized strategies' performance, especially when other agents’ state information might be inaccurate, we evaluate the proposed approach in scenarios where players only have access to noisy state estimates of other players.
Players make decisions separately with perfect state information about themselves and noisy state information about other players. 
We run a Monte Carlo study under zero-mean isotropic Gaussian noise with~50 different variance levels starting from zero. 
\tacinitial{
The reference trajectory in \cref{fig:formation-game} spans a square area measuring 60 meters $\times$ 60 meters. 
The maximum standard deviation of the added observation noise is around 45 meters (corresponding to a maximum per-dimension variance of 2000), which is about 75\% of the reference trajectory length/height.} 
We compare standard Nash equilibrium strategies against the regularized strategies by our approach with~100 different regularization levels. For each combination of a noise and a regularization level, we run both approaches starting from~100 random players' positions uniformly sampled from a square measured 20 meters~$\times$ 20 meters and centred at the origin. Hence, the experiment includes~$5\times 10^5$ comparisons in total.

\Cref{fig:formation-game-cost} shows the cost performance of the regularized strategies subtracted by standard Nash equilibrium costs and averaged over~100 random players' initial positions, where a negative value indicates that the proposed approach yields better performance in this setting. 

Recall that player~1's cost is not coupled with the other players and we only regularize the off-diagonal blocks in the policy matrices, corresponding to inter-agent dependencies. 
As shown in~\cref{fig:formation-game-cost}, player~1's performance is invariant with different noise levels and almost invariant with different regularization levels.
The cost of player~1 has a tiny increase with higher regularization.
This is because the block $(1,1)$ of the policy matrices is slightly biased by the regularization applied to neighborhood blocks, which we shall explain in detail in Appendix. 
Nonetheless, the cost change is only up to 0.35\%.

Players~2 and~3 show the same trend in their performance with different noise and regularization levels. 
In columns~2 and~3 of~\cref{fig:formation-game-cost}, the parts above the orange zero level set line indicate trials where the proposed approach performs better. 
On a high level, the regularized strategies computed by the proposed approach outperform the standard Nash equilibrium strategies for the majority of the trials (around 82\%). 
An important observation is: \emph{For any positive (even if small) noise level, there exists an interval of regularization levels where the sparse policies perform better than the standard Nash equilibrium policies; the optimal regularization level is always strictly positive for any nonzero noise level.} 
As the noise level increases, the interval where the regularized policies perform better also grows. 
Above noise level around 500, any regularized policies computed by the proposed approach yield lower costs than the standard Nash equilibrium. 

\emph{Therefore, the results above validate the claim \textbf{C4}.}

\noindent\textbf{Qualitative Results.} \Cref{fig:formation-game-qualitative} shows the resulting trajectories from standard Nash equilibrium strategies compared with the proposed approach with different regularization levels. 
Under noise-corrupted state estimates, standard Nash equilibrium strategies greedily optimize players 2 and 3's costs and yield noisy, high-cost trajectories. 
By contrast, with higher regularization, the proposed approach regularizes the solutions more and forces the strategies to be less dependent on other players' states. 
As a result, players 2 and 3 move in smaller sinusoidal shapes and their motion becomes less sensitive to other players' noisy states. 
In the extreme high-regularization case~$\lambda = 15$, players 2 and 3's strategies become completely independent of other players and yield smooth sinusoidal trajectories. 
Among the regularization levels,~$\lambda = 5$ gives the lowest costs for players 2 and 3 in this setting.

\section{Conclusion}

We have presented a regularized \ac{dp} approach for finding sparse feedback policies in noncooperative games. 
The regularized policies selectively depend on a subset of agents' states, thereby reducing the communication and sensing resources required for execution. 
The proposed approach encodes structured sparsity into \ac{lq} game solutions via solving a convex adaptive group Lasso problem at each \ac{dp} iteration. 
This approach is further extended to general non-\ac{lq} games via iterative \ac{lq} approximations. 

For \ac{lq} games, we establish an upper bound on the single-stage difference between the regularized and standard Nash equilibrium policy matrices. 
Thus equipped, we prove the asymptotic convergence of the regularized solution to a neighborhood of the Nash equilibrium across all game stages.

Extensive results in multi-robot navigation  
tasks validate the efficacy of the proposed approach in producing sparse feedback policies at varying regularization levels and demonstrate its scalability to games involving multiple players. 
Monte Carlo studies in a multi-agent formation game also show that when interacting agents with coupled costs have noisy observations of other agents' states, 
it is always better to employ some (nonzero) degree of regularization. 
In this case, the regularized policies consistently outperform standard Nash equilibrium solutions by selectively ignoring information that may be inaccurate.

Future work could explore the optimal selection of regularization strengths across different game stages. Additionally, applying the proposed regularization scheme to scenarios involving more complex uncertainties beyond observation noise is also an interesting direction. 
Lastly, a deeper investigation into the observation presented in \cref{sec:multi-agent-navigation}---where our approach yields distinct local solutions at different regularization levels in non-\ac{lq} games—would be of interest.

\balance

\section*{Acknowledgment}

We thank Yue Yu for his helpful discussions about convergence analysis.

\section*{References}
\bibliographystyle{IEEEtran}
\bibliography{IEEEabrv,main}

\appendix
\section{Appendix}
\label{appendix}

\begin{customproof}[Proof of \cref{lemma:one-step-bound}]\label{sec:proof-lemma-1}
We rewrite the unregularized system of equations in \cref{eq:system-equations} as an optimization problem:
\begin{equation}\label{eq:original-problem}
    \min_{\Pt} \underbrace{\frac{1}{2} \| \St \Pt - \Yt \|_F^2}_{h(\Pt)},
\end{equation}
and we have the regularized problem in~\cref{eq:reg-problem}:
\begin{equation}\label{eq:reg-problem-appendix}
    \min_{\Pht} \fhat(\Pht) = \underbrace{\frac{1}{2} \| \St \Pht - \Yt \|_F^2}_{h(\Pht)} + \underbrace{\sum_{i,j} \lambda_{i,j} \|\Pht^i[j]\|_F}_{\psi(\Pht)}, 
\end{equation}
where~$\Pht^i[j] \in \R^{n^i \times m^j}$ denotes the $(i,j)$ block of the~$\Pht$ matrix.
We recall the dimensions of players' states and controls~$x_t^i \in \R^{m^i}, u_t^i \in \R^{n^i}$, and~$m = \sum_{i \in [\numplayers]}m^i$,~$n = \sum_{i \in [\numplayers]}n^i$. 
Hence, we have~$\St \in \R^{ n \times  n}$, and $\Pt,\Pht \in \R^{n \times m}$.

First, we examine the optimality conditions for both problems and characterize their solutions. 
Then, we bound the difference between the two solutions. 

\subsubsection*{Optimality conditions}

The unregularized problem in~\cref{eq:original-problem} is a convex optimization problem with a differentiable objective function, whose necessary and sufficient optimality condition~\cite[Ch.4]{boyd2004convex} is:
\begin{multline}\label{eq:original-solution}
    \nabla_{\Pt} h(\Pt) = 0 \Rightarrow \Mt \Pt - \St^\top \Yt = 0
    \Rightarrow \Pt = \Mt^{-1}\St^\top \Yt,
\end{multline}
where we define~$\Mt = \St^\top \St$ and recall that $\St$ is full rank by \cref{assumption:unique-nash}. 

The objective of the regularized problem in~\cref{eq:reg-problem-appendix} is nonsmooth but convex. 
Since~$h$ is convex and differentiable and~$\psi$ is convex and finite-valued, we have the necessary and sufficient optimality condition~\cite[Thm 8.17]{rechtwright}:
\begin{multline*}
    0 \in \nabla_{\Pht} h(\Pht) + \partial_{\Pht}\psi(\Pht) \\= \Mt \Pht - \St^\top \Yt +
    \partial_{\Pht}(\sum_{i,j} \lambda_{i,j} \|\Pht^i[j]\|_F),
\end{multline*}
where~$\partial(\cdot)$ denotes the subdifferential set. 
Since~$\psi$ is convex and finite-valued~\cite[Thm 8.11]{rechtwright}, we have:
\begin{equation*}
     \partial_{\Pht}\psi = \partial_{\Pht}(\sum_{i,j} \lambda_{i,j} \|\Pht^i[j]\|_F) = \sum_{i,j} \lambda_{i,j} \partial_{\Pht} (\|\Pht^i[j]\|_F).
\end{equation*}
Therefore, we obtain the optimality condition:
\begin{equation}\label{eq:reg-optimality}
     0 \in  \Mt \Pht - \St^\top \Yt + \sum_{i,j} \lambda_{i,j} \partial_{\Pht} (\|\Pht^i[j]\|_F).
\end{equation}

The subdifferential for~$\|\Pht^i[j]\|_F$ is given by:
\small
\begin{equation}\label{eq:reg-subdiff}
    \partial_{\Pht^i[j]}(\|\Pht^i[j]\|_F) = 
    \begin{cases}
    
        \frac{\Pht^{i}[j]}{\|\Pht^{i}[j]\|_F}, \Pht^{i}[j] \neq 0\\

        \{z \given z \in \R^{n^i \times m^j}, \|z\|_F \leq 1\}, \Pht^{i}[j] = 0.
        
    \end{cases}
\end{equation}
\normalsize
Therefore, elements in the subdifferential~$\partial_{\Pht}(\|\Pht^i[j]\|_F)$ are matrices that have entries shown in~\cref{eq:reg-subdiff} for block $(i,j)$ and zeros elsewhere.

For a solution~$ \Pht$ that satisfies the optimality condition in~\cref{eq:reg-optimality}, we must be able to find a particular set of subgradients $\theta^*_{1,1}$, $\theta^*_{1,2}$, $\hdots$, $\theta^*_{\numplayers,\numplayers}$ from the subdifferential sets such that:
\begin{equation*}
    0 =  \Mt \Pht - \St^\top \Yt + \sum_{i,j} \lambda_{i,j} \theta_{i,j}^*,
\end{equation*}
where~$ \theta^*_{i,j} \in \R^{n \times m}, \theta^*_{i,j} \in \partial_{\Pht}(\|\Pht^i[j]\|_F)$.
Such a point specifies a solution for the regularized problem, i.e.
\begin{equation}\label{eq:reg-solution}
    \Pht = \Mt^{-1}\St^\top\Yt - \sum_{i,j} \lambda_{i,j} \Mt^{-1}\theta_{i,j}^*.
 \end{equation}

\subsubsection*{Bounding the solution difference}

Having characterized solutions to the two problems in~\cref{eq:original-problem,eq:reg-problem-appendix}, we summarize~\cref{eq:original-solution} and~\cref{eq:reg-solution} to compute the difference between the solutions:
\begin{multline}
    \|\Delta \Pt \|_F = \|\Pht - \Pt\|_F = \norm{\sum_{i,j} \lambda_{i,j} \Mt^{-1}\theta_{i,j}^*}_F\\
    \leq \sum_{i,j} \lambda_{i,j} \norm{\Mt^{-1}\theta_{i,j}^*}_F.
\end{multline}
Next, we split the terms in~$\sum_{i,j} \lambda_{i,j} \norm{\Mt^{-1}\theta_{i,j}^*}_F$ into two cases based on~\cref{eq:reg-subdiff}.\\

\emph{Case 1.} For blocks $(i,j)$ such that~$\Pht^{i}[j] \neq 0$, we have that~$\normf{\theta^*_{i,j}} = 1$. 
We denote a vectorization operator as $\mathrm{vec(\cdot)}$. 
We note that we can rewrite the multiplication~$\Mt^{-1}\theta_{i,j}^*$ in a matrix-vector form such that~$\norm{\Tilde{\Mt}^{-1}\thetat}_2 = \norm{\Mt^{-1}\theta_{i,j}^*}_F$, where the matrix $\Tilde{\Mt}^{-1} := \mathrm{blogdiag}(\Mt^{-1}, \hdots,\inv{\Mt}) \in \R^{mn \times mn}$ is the block diagonal concatenation of $m$ copies of~$\inv{\Mt}$, and~$\thetat := \vect{(\theta^*_{i,j})} \in \R^{mn}$. 

From the definition of the matrix spectral norm, we have:
\begin{equation}
    \norm{\Tilde{\Mt}^{-1}}_2 = \sup_{ \norm{ \thetat }_2 = 1} \norm{\Tilde{\Mt}^{-1}\thetat}_2
\end{equation}
Hence,
\begin{equation}
\begin{aligned}
     \lambda_{i,j} \norm{\Mt^{-1}\theta^*_{i,j}}_F &= \lambda_{i,j} \norm{\Tilde{\Mt}^{-1}\thetat}_2 \\
     &\leq \lambda_{i,j} \norm{\Tilde{\Mt}^{-1}}_2 = \lambda_{i,j} \sigma_{\mathrm{max}}(\inv{\Tilde{\Mt}}),
\end{aligned}
\end{equation}
where~$\sigma$ denotes singular values. We note that, since~$\Tilde{\Mt}$ has a block-diagonal structure, it is straightforward to verify that~$\sigma_{\mathrm{max}}(\inv{\Tilde{\Mt}}) = \sigma_{\mathrm{max}}(\inv{\Mt})$.\\

\emph{Case 2.} For blocks $(i,j)$ such that~$\Pht^{i}[j] = 0$, since we know $\norm{ \thetat }_2 \leq 1$, we have:
\begin{equation}
\begin{aligned}
    \norm{\Mt^{-1}\theta^*_{i,j}}_F = \norm{\Tilde{\Mt}^{-1}\thetat}_2 &\leq \sup_{\norm{ \thetat }_2 \leq 1} \norm{\Tilde{\Mt}^{-1}\thetat}_2\\
    &\leq  \sup_{\norm{ \thetat }_2 = 1} \norm{\Tilde{\Mt}^{-1}\thetat}_2\\
    &= \norm{\Tilde{\Mt}^{-1}}_2\\
    &= \sigma_{\mathrm{max}}(\inv{\Tilde{\Mt}})\\
    &= \sigma_{\mathrm{max}}(\inv{\Mt}).
\end{aligned}
\end{equation}\\

Recall that we require the original system of equations in \cref{eq:system-equations} to have a unique solution, \ie,~$\St$ has full rank. 
Therefore,~$\Mt$ has positive singular values and it follows that~$\maxsv{\inv{\Mt}} = \frac{1}{\minsv{\Mt}} = \frac{1}{\sigma_{\mathrm{min}}^2(\St)}$. 
Hence, summarizing the two cases,
\begin{equation}\label{eq:error-bound-T}
    \normf{\Delta \Pt} \leq \frac{\sum_{i,j \in [\numplayers]} \lambda_{i,j}}{\sigma_{\mathrm{min}}^2(\St)}
\end{equation}
bounds the difference between the regularized and unregularized solutions at a single time step $t$.
\end{customproof}

\begin{customproof}[Proof of \cref{theorem:iss}]\label{sec:proof-theorem-2}
\tacinitial{
We first define the dynamical system $\StateRiccatiAllReg_t = \DynRiccati(\StateRiccatiAllReg_{t+1}, \delta_t)$ studied in \cref{theorem:iss} and then prove \cref{theorem:iss}.
}

\subsubsection*{Definition of the Dynamical System}


For all $i \in [\numplayers]$, we have algebraic coupled Riccati equations in \cref{eq:inf-horizon-riccati} for infinite-horizon \ac{lq} game:
\begin{equation}\label{eq:inf-riccati}
    Z^{i*} = Q^i + \sum_{j=1}^\numplayers P^{j*\top} R^{ij} P^{j*} + F^{*\top} Z^{i*}F^*.
\end{equation}
For finite-horizon \ac{lq} games, we have unregularized coupled Riccati recursion~\cref{eq:value-func}:
\begin{equation}\label{eq:finite-unreg-riccati}
    Z_t^i = Q^i + \sum_{j=1}^\numplayers P_t^{j\top} R^{ij} P_t^{j} + F_t^\top Z^i_{t+1}F_t,
\end{equation}
and regularized coupled Riccati recursion in~\cref{eq:regularized-riccati}:
\begin{equation}\label{eq:finite-reg-riccati}
    \hat{Z}_t^i = Q^i + \sum_{j=1}^\numplayers \Pht^{j\top} R^{ij} \Pht^{j} + \Fht^\top \hat{Z}^i_{t+1}\Fht,
\end{equation}
where:
\begin{equation*}
    \Pht^j = P^j_t + \Delta P^j_t,\quad \Fht = F_t - \sum_{j = 1 }^\numplayers B^j\Delta P^j_t.
\end{equation*}

In order to study the convergence of the regularized coupled Riccati recursion, we analyze the asymptotic difference between~\cref{eq:finite-reg-riccati} and~\cref{eq:inf-riccati}, for all $i \in [\numplayers]$:

\medskip
\medskip
\medskip

\small
\begin{fleqn}[0pt]
\begin{equation*}
    \begin{aligned}
        &\StateRiccatiReg_t := \hat{Z}_t^i - Z^{i*} = \sum_{j=1}^\numplayers \Pht^{j\top} R^{ij} \Pht^{j} 
        + \Fht^\top \hat{Z}^i_{t+1}\Fht -\\ 
        &\quad \sum_{j=1}^\numplayers P^{j*\top} R^{ij} P^{j*} - F^{*\top} Z^{i*}F^* \\
    \end{aligned}
\end{equation*}
\normalsize
\small
\begin{equation}\label{eq:split-two-parts}
    \begin{aligned}
        &= \sum_{j=1}^\numplayers (P_t^j + \Delta P_t^j)^\top R^{ij} (P_t^j + \Delta P_t^j) 
        + (F_t - \sum_{j=1}^\numplayers B^j\Delta P_t^j)^\top \cdot \\
        &\quad  \hat{Z}_{t+1}^i(F_t - \sum_{k=1}^\numplayers B^k\Delta P_t^k) 
        - \sum_{j=1}^\numplayers P^{j*\top} R^{ij} P^{j*} 
        - F^{*\top} Z^{i*}F^* \\
        &= \textcolor{black}{\sum_{j=1}^\numplayers P_t^{j^\top} R^{ij} P_t^j  
        + F_t^\top \hat{Z}_{t+1}^i F_t 
        - \sum_{j=1}^\numplayers P^{j*\top} R^{ij} P^{j*} 
        -} \\
        & \quad \textcolor{black}{F^{*\top} Z^{i*}F^*} \\
        &\textcolor{black}{\quad + \sum_{j=1}^\numplayers \Delta P_t^{j\top}R^{ij}P_t^j 
        + \sum_{j=1}^\numplayers \Delta P_t^{j\top}R^{ij}\Delta P_t^j 
        + \sum_{j=1}^\numplayers P_t^{j\top}R^{ij}\Delta P_t^j} \\
        &\textcolor{black}{\quad - \sum_{j=1}^\numplayers \Delta P_t^{j\top} B^{j\top} \hat{Z}_{t+1}^i F_t 
        - F_t^\top \hat{Z}_{t+1}^i \sum_{k=1}^\numplayers B^k \Delta P_t^{k}} \\
        &\textcolor{black}{\quad + \sum_{j=1}^\numplayers \Delta P_t^{j\top} B^{j\top} \hat{Z}_{t+1}^i 
        \sum_{k=1}^\numplayers B^k \Delta P_t^k}\\
        &= \textcolor{black}{\RegRiccatiAutonomous^i(\hat{Z}_{t+1})} + \textcolor{black}{\RegRiccatiDisturb^i(\hat{Z}_{t+1}, \Delta P_t)}.
    \end{aligned}
\end{equation}
\end{fleqn}
\normalsize
Hence:
\begin{equation}\label{eq:reg-dynamical-sys}
    \StateRiccatiReg_t = \hat{Z}_t^i - Z^{i*} = \textcolor{black}{\RegRiccatiAutonomous^i(\hat{Z}_{t+1})} + \textcolor{black}{\RegRiccatiDisturb^i(\hat{Z}_{t+1}, \Delta P_t)}.
\end{equation}
Aggregating all the players $\StateRiccatiAllReg_t := [\StateRiccatiAllReg^{1\top}_t, \dots, \StateRiccatiAllReg^{\numplayers\top}_t]^\top$, we have:
\begin{equation}\label{eq:reg-joint-dynamical-sys}
    \StateRiccatiAllReg_t = \hat{Z}_t - Z^* = \textcolor{black}{\RegRiccatiAutonomous(\hat{Z}_{t+1})} + \textcolor{black}{\RegRiccatiDisturb(\hat{Z}_{t+1}, \Delta P_t)}.
\end{equation}
We define the dynamical system in \cref{eq:reg-joint-dynamical-sys} as:
\begin{equation}\label{eq:def-reg-joint-dynamical-sys}
    \StateRiccatiAllReg_t = \DynRiccati(\StateRiccatiAllReg_{t+1}, \delta_t),
\end{equation}
where we denote regularization difference $\Delta P_t$ as disturbance $\delta_t$. The system in \cref{eq:def-reg-joint-dynamical-sys} is the dynamical system we referred to in \cref{theorem:iss}.

\subsubsection*{Proof of \cref{theorem:iss}}

In \cref{eq:def-reg-joint-dynamical-sys}, 
the system dynamics $\DynRiccati(\StateRiccatiAllReg_{t+1}, \delta_t)$ consist of the nominal part $\textcolor{black}{\RegRiccatiAutonomous(\hat{Z}_{t+1})}$ and the part which directly depends upon  disturbances~$\textcolor{black}{\RegRiccatiDisturb(\hat{Z}_{t+1}, \Delta P_t)}$, as expanded in \cref{eq:split-two-parts}. 
When there is no regularization, \ie, $\Delta P_t = 0$ and~$\hat{Z}_t = Z_t, \forall t$, we have $\textcolor{black}{\RegRiccatiDisturb(\hat{Z}_{t+1}, \Delta P_t)} = 0$ and  
the system $ \StateRiccatiAllReg_t = \DynRiccati(\StateRiccatiAllReg_{t+1}, \delta_t)$ reduces to $\StateRiccatiAll_t = \DynRiccati(\StateRiccatiAll_{t+1}, 0)$, which only has the nominal part and is exactly the difference between \cref{eq:finite-unreg-riccati} and \cref{eq:inf-riccati} aggregated over all the players. 
That is, at zero regularization, the dynamical system in \cref{eq:def-reg-joint-dynamical-sys} reduces to the difference between the unregularized finite-horizon Riccati and infinite-horizon Riccati equations. 

By \cref{assumption:convergence-riccati}, the dynamical system in \cref{eq:def-reg-joint-dynamical-sys} with zero regularization $\StateRiccatiAll_t = \DynRiccati(\StateRiccatiAll_{t+1}, 0) = \textcolor{black}{\RegRiccatiAutonomous(Z_{t+1})}$ is \acl{las} at the origin. 
Furthermore, the system in \cref{eq:def-reg-joint-dynamical-sys} is continuous in both value matrix~$\hat{Z}_{t+1}$ and disturbance~$\Delta P_t$, and therefore in $\StateRiccatiAllReg_{t+1}$ and $\delta_t$. 
Hence, the system in \cref{eq:def-reg-joint-dynamical-sys} is locally \ac{iss} with respect to the regularization~\cite[Lemma 2.2]{jiang2004nonlinear}. 

We recall the definitions of class~$\mathcal{K}$ and class~$\mathcal{KL}$ functions~\cite{khalil2002nonlinear}:
\begin{definition} 

A continuous function $\BallRiccati: \R_{\geq 0} \rightarrow \R_{\geq 0}$ belongs to class~$\mathcal{K}$ if it is strictly increasing with $\BallRiccati(0) = 0$.
A continuous function $\RiccatiTransient: \R_{\geq 0} \times \R_{\geq 0} \rightarrow \R_{\geq 0}$ belongs to class~$\mathcal{KL}$ if $\RiccatiTransient(\cdot, k)$ is of class $\mathcal{K}$ for each fixed $k$, and $\RiccatiTransient(r, \cdot)$ is decreasing for each fixed $r$, and $\lim_{k \rightarrow \infty} \RiccatiTransient(r, k) = 0$.

\end{definition}

From the definition of local input-to-state stability~\cite[Definition 2.1]{jiang2004nonlinear}, there exist a class~$\mathcal{KL}$ function~$\RiccatiTransient$ and a class $\mathcal{K}$ function $\BallRiccati$ such that: 
\begin{equation}\label{eq:iss-appendix}
    \normf{\StateRiccatiAllReg_t} \leq \RiccatiTransient(\normf{\StateRiccatiAllReg_{\horizon + 1}}, \horizon - t) + \BallRiccati(\norm{\delta_t}_{\infty}), \forall t \in (-\infty, \horizon],
\end{equation}
for disturbances such that~$\sup_{t \leq\tau\leq \horizon}\normf{\delta_\tau} = \norm{\delta_t}_{\infty} < \delta, \delta \in \R_{>0}$, and $\|\StateRiccatiAllReg_{\horizon + 1}\|_F \leq z, z \in \R_{>0}$.
\end{customproof}

\begin{customproof}[Proof of \cref{corollary:iss}]\label{sec:proof-corollary-3}
In \cref{eq:iss-appendix}, since~$\RiccatiTransient(\normf{\StateRiccatiAllReg_{\horizon + 1}}, \horizon - t)$ decays to 0 as $t \rightarrow -\infty$, for every $\epsilon>0$ there exists $t^\star\in(-\infty,T]$ such that if $t\in(-\infty,t^\star]$ then
\begin{equation}\label{eq:iss-ball}
    \normf{\StateRiccatiAllReg_t} \leq \frac{\epsilon}{2} +\BallRiccati(\norm{\delta_t}_{\infty}),~\forall t\in(-\infty,t^\star].
\end{equation}
In addition, because $\BallRiccati(\norm{\delta_t}_{\infty})$ is class $\mathcal{K}$, for an arbitrary ball with radius $\epsilon > 0$, there exists a positive constant $ \xi_1 > 0$ such that the following relation holds:
\begin{equation}
\label{eq:condition-disturbance}
    \norm{\delta_t}_{\infty} \leq \xi_1 \implies
    \BallRiccati(\norm{\delta_t}_{\infty})\le\frac{\epsilon}{2}.
\end{equation}
Therefore, combining \cref{eq:iss-ball} and \cref{eq:condition-disturbance} \revised{and given that the conditions in \cref{theorem:iss} hold}, for every $\epsilon>0$ there exists $t^\star\in(-\infty,T]$ and $\xi_1>0$ such that if $\norm{\delta_t}_{\infty} \leq \xi_1$ then
\begin{equation}
\label{eq:condition-disturbance2}
    \normf{\StateRiccatiAllReg_t} \leq \epsilon,~ t\in(-\infty,t^\star].
\end{equation}
\revised{That is, for small disturbances $\delta_t$, $\StateRiccatiAllReg_t$ will eventually enter and stay inside the ball of radius $\epsilon$.}

\cref{eq:condition-disturbance2} is true given a condition on $\norm{\delta_t}_{\infty}$, which we now need to relate to the  regularization constants~$\lambda_{i,j}$. To that end, recall that the regularized $\hat{S}_t$ takes the form:
\small
\begin{multline}\label{eq:finite-horizon-S}
\begin{aligned}
        \hat{S}_t = \resizebox{0.83\linewidth}{!}{$
\begin{bmatrix}
        R^{11}+B^{1\top}\hat{Z}_{t+1}^1B^1 &  \hdots & B^{1\top}\hat{Z}_{t+1}^1B^N\\
        B^{2\top}\hat{Z}_{t+1}^2B^1 &  \hdots & 
        B^{2\top}\hat{Z}_{t+1}^2B^N\\
        \vdots &  \ddots & \vdots \\
        B^{N\top}\hat{Z}_{t+1}^NB^1 &  \hdots & R^{NN} + B^{N\top}\hat{Z}_{t+1}^NB^N\\
    \end{bmatrix}.
    $}
\end{aligned}
\end{multline}
\normalsize
For infinite-horizon \ac{lq} games, taking the limit $t \rightarrow -\infty$, we have the infinite-horizon counterpart of $S_t$ as follows:
\small
\begin{multline}\label{eq:inf-horizon-S}
\begin{aligned}
        S^* = 
\begin{bmatrix}
        R^{11}+B^{1\top}Z^{1*}B^1 &  \hdots & B^{1\top}
        Z^{1*}B^N\\
        B^{2\top}Z^{2*}B^1 &  \hdots & 
        B^{2\top}Z^{2*}B^N\\
        \vdots & \ddots & \vdots \\
        B^{N\top}Z^{N*}B^1 & \hdots & R^{NN} + B^{N\top}
        Z^{N*}B^N\\
    \end{bmatrix}.
\end{aligned}
\end{multline}
\normalsize
By \cref{assumption:unique-nash}, the unregularized $S_t$ in \cref{eq:system-equations} and $S^*$ in \cref{eq:inf-horizon-S} are of full rank. Therefore, $\inf_{t} \minsv{S_t} = v> 0$. In addition, we observe that when~$\sum_{i,j \in [\numplayers]} \lambda_{i,j} \rightarrow 0$, we have $\hat{S}_t \rightarrow S_t$ and $\hat{Z}_t \rightarrow Z_t$ pointwise in $t$. By the continuity of the singular values, this implies that for every $t^\star\in(-\infty,T]$ there exists $\xi_2>0$ such that if $\sum_{i,j \in [\numplayers]} \lambda_{i,j}\le \xi_2$ then $\minsv{\hat{S}_t} \ge v/2$\footnote{Note that here $v/2$ is just one particular choice, any positive number smaller than $v$ suffices.} for all $t\in[t^\star,T]$. In addition, it implies that
$\normf{\StateRiccatiAllReg_{t^\star}}\le\epsilon$ if $t^\star$ is chosen to be sufficiently small,  because 
$Z_t\rightarrow Z^*$ as $t\rightarrow-\infty$. 
\revised{In other words, for small disturbances,  $\minsv{\hat{S}_t}$ will remain uniformly away from zero for a given amount $t\in[t^\star,T]$, and $\hat{Z}_t$ will enter a ball of radius $\epsilon$ around $Z^*$ at $t^\star$. It then remains to show that $\hat{Z}_t$ will also stay in that ball for all $t\le t^\star$.}

Given that $\normf{\StateRiccatiAllReg_{t^\star}}\le\epsilon$, we can subsequently prove that $\normf{\StateRiccatiAllReg_{t}}\le\epsilon$ for all $t\in(-\infty,t^\star]$ by induction. In particular, from \cref{lemma:one-step-bound}, if we require 
\begin{equation}\label{eq:sum_l_xi1}
    \frac{\sum_{i,j \in [\numplayers]} \lambda_{i,j}}{(\frac{\MinSingularValue}{2})^2} \leq \xi_1,
\end{equation}
then we guarantee $\eqref{eq:condition-disturbance}$ holds at time $t=t^\star$, and thus $\normf{\StateRiccatiAllReg_{t^\star-1}}\le \epsilon$ from \cref{eq:iss-ball}. Choosing $\epsilon>0$ small enough so that $\minsv{\hat{S}_t} \ge v/2$ for any $\normf{\StateRiccatiAllReg_t}\le\epsilon$, it follows that $\minsv{\hat{S}_{t^\star-1}} \ge v/2$, hence also $\normf{\StateRiccatiAllReg_{t^\star-2}}\le \epsilon$ by \cref{eq:sum_l_xi1} and \cref{lemma:one-step-bound}. We can follow this argument recursively and show that $\normf{\StateRiccatiAllReg_{t}}\le \epsilon$ for all $t\in(-\infty,t^\star]$, and thus
\begin{equation}
     \limsup_{t \rightarrow -\infty} \normf{\StateRiccatiAllReg_t} \leq \epsilon.
\end{equation}
To conclude, recall that the above results hold if we choose $\xi$ in \cref{corollary:iss} to be $\xi = \mathrm{min}((\frac{\MinSingularValue}{2})^2\xi_1, \xi_2)$. 
\end{customproof}

\begin{appendixsec}[Transformation of Group Lasso Problem to a Conic Program]\label{sec:appendix-group-lasso-transform}
The group Lasso problem in \cref{eq:reg-problem} can be reformulated as a conic program, as described below. 
Observe that \cref{eq:reg-problem} can be rewritten as:
\begin{equation}\label{eq:appendix classical formulation of group lasso}
        \min_{\hat{P}_t}  \frac{1}{2}\|(I_N \otimes S_t ) \textrm{vec}(\hat{P}_t) - \textrm{vec}(Y_t)\|_2^2 \\
        + \sum_{i,j} \lambda_{i,j} \|\textrm{vec}(\Pht^i[j])\|_2,
\end{equation}
where $\otimes$ is the Kronecker product. 
Let ${s_{i,j}\in \R}$ be a slack variable. We define $s:= \{s_{i,j}, \forall i, j\in [N]\}$. By using the epigraph trick \cite{boyd2004convex}, we can rewrite (\ref{eq:appendix classical formulation of group lasso}) as a conic program with quadratic cost: %
\begin{equation}
    \begin{aligned}
        \min_{\hat{P}_t, s} & \frac{1}{2} \|(I_N\otimes S_t) \textrm{vec}(\hat{P}_t) - \textrm{vec}(Y_t)\|_2^2 + \sum_{i,j} \lambda_{i,j} s_{i,j}\\
        \textrm{s.t. }& \qquad \|\textrm{vec}(\Pht^i[j])\|_2 \le s_{i,j},\ \ \forall i,j\in[N] %
    \end{aligned}
\end{equation}
and the constraint is equivalent to 
\begin{equation}
    \begin{bmatrix}
        s_{i,j} \\ \textrm{vec}(\hat{P}_t^i[j])
    \end{bmatrix} \in K_{\textrm{soc}},\ \ i,j\in[N]
\end{equation}
where $K_{\textrm{soc}}$ is a second-order cone \cite{boyd2004convex}, defined as $K_{\textrm{soc}}:=\{(p,q):p\in\mathbb{R}^{m_i\times n_i} ,q\in\mathbb{R}, \|p\|_2\le q\}$. 
The conic program with quadratic objective can be solved using off-the-shelf solvers, \eg, \cite{Clarabel_2024}. 
\end{appendixsec}

\begin{appendixsec}[Game Cost Functions]\label{sec:game-cost}
\tacinitial{The cost functions used in the two games in \cref{sec:results} are given below.}

\subsubsection*{Navigation Game}
In the navigation game in \cref{sec:multi-agent-navigation}, each player $i$ seeks to minimize a cost function, which is the sum of running costs for each time step $t$ over the horizon $T$. 
The running cost at time $t$ is a sum of the following terms:

\noindent (\romannumeral 1) goal reaching: 
\begin{equation*}
     \mathbbm{1}\{t > t_{\mathrm{active}}\} \cdot \frac{1}{2}(x_t^i - x^i_{\mathrm{goal}})^\top 
     \resizebox{0.28\linewidth}{!}{$
    \begin{bmatrix}
        300 &  &  &  \\
        & 300 & &  \\
         &  & 300 &  \\
        &  &  & 0\\
    \end{bmatrix}
    $}
    (x_t^i - x^i_{\mathrm{goal}}) 
\end{equation*}

\noindent (\romannumeral 2) velocity: penalizing large velocity
\begin{equation*}
    \frac{1}{2}30(v_t^i)^2 + 50 \cdot \mathrm{soft\_constraint}(v_t^i, -0.05, 2),
\end{equation*}
where the $\mathrm{soft\_constraint}(r, \mathrm{min}, \mathrm{max})$ is a function that takes the form:
\begin{equation*}
    \mathrm{soft\_constraint}(r, \mathrm{min}, \mathrm{max}) = 
    \begin{cases}
    
        (r - \mathrm{min})^2, r < \mathrm{min}\\

        (r - \mathrm{max})^2, r > \mathrm{max}\\

        0, \mathrm{otherwise}.
        
    \end{cases}
\end{equation*}

\noindent (\romannumeral 3) proximity avoiding: 
\begin{equation*}
     \sum_{j \in [\numplayers], j \neq i} 50 \cdot \mathrm{max}(0, d_{\mathrm{min}} - \norm{p_t^i - p_t^j}_2)^2
\end{equation*}

\noindent (\romannumeral 4) control: 
\begin{equation*}
\begin{aligned}
    \frac{1}{2}u_t^{i\top}
    \begin{bmatrix}
        10 &    \\
        & 10  \\
    \end{bmatrix} u_t^i + 50\cdot \mathrm{soft\_constraint}(\omega_t^i, -\frac{\pi}{18}, \frac{\pi}{18}) \\
    + 50\cdot \mathrm{soft\_constraint}(a_t^i, -9.81, 9.81) 
\end{aligned}
\end{equation*}

In the terms above, we recall $p_t^i$ denotes player $i$'s position at time~$t$. 
$\mathbbm{1} \{\cdot\}$ denotes an indicator function, which takes value of 1 if the statement holds true and 0 otherwise. 
Moreover, $t_{\mathrm{active}} = 0.99 t_{\mathrm{total}}$ is a time step after which the goal-reaching cost becomes active. The total time $t_{\mathrm{total}} = 15 \si{\second}$ and the horizon $T = 150$ in this example. The minimum distance that activates the proximity cost $d_{\mathrm{min}} = 0.5 \si{\meter}$. 

\subsubsection*{Formation Game}

In the formation game in \cref{sec:formation-game}, player~1 tracks a prespecified trajectory, whose cost is independent of other players:
\begin{equation*}
    J^1 = \sum_{t \in [T]} \frac{1}{2}1000\norm{p_t^1 - p_{t,\mathrm{goal}}}_2^2+ \frac{1}{2}\norm{u_t^1}_2^2.
\end{equation*}
The target trajectory is a sinusoidal trajectory in \cref{fig:formation-game}. At time $t$, the target position is $[30\cdot\cos{(t \cdot \Delta T)}, 30\cdot\sin{(3t \cdot \Delta T)}]^\top$, with $\Delta T  = 0.05 \si{\second}$ being the time discretization interval.

Players 2 and 3 are tasked to maintain a formation relative to the other two players. For each player $i \in \{2, 3\}$, the cost function is:
\begin{equation*}
\begin{aligned}
    J^i = \sum_{t \in [T]}\sum_{j\in [\numplayers], j \neq i} \frac{1}{2} 1000 
    \norm{ p_t^i - (p_t^j - \Delta p_{t, \mathrm{rel}}^{i,j})}_2^2
    + \frac{1}{2}\norm{u_t^{i}}_2^2,  
\end{aligned}
\end{equation*}
where $\Delta p_{t, \mathrm{rel}}^{i,j}$ is the target relative position between players $i$ and $j$ at time $t$, which determines the formation. In this example, the target formation is always an isosceles triangle with a width of $4 \si{\meter}$ and height of $2 \si{\meter}$. 
\end{appendixsec}

\begin{appendixsec}[Explanation of Player 1's Costs in \cref{fig:formation-game-cost}]\label{sec:appendix-explanation-p1-cost}
In \cref{fig:formation-game-cost}, player 1's cost is independent of the other players, and we do not add a regularizer for block (1,1) in the policy matrix $P_t$. 
However, the cost of player 1 still increases by up to 0.35\% as the regularization level rises.

This result may initially seem counterintuitive, but it can be explained by \cref{eq:reg-solution}. 
At the terminal stage, the single-stage difference between the unregularized policy $P_t$ and regularized policy $\Pht$ is given by $- \sum_{i,j} \lambda_{i,j} \Mt^{-1}\theta_{i,j}^*$. 
The matrix $\Mt^{-1}$ is generally dense in our experiments. 
Therefore, although $\lambda_{1,1}$ is zero, the non-zero values of $\lambda_{2,1}$ and $\lambda_{3,1}$ mean that $- \sum_{i \in \{2,3\}} \lambda_{i,1} \Mt^{-1}\theta_{i,1}^*$ may still introduce a small bias to block (1,1) of the policy matrix $\Pht$.
This effect can similarly occur at earlier stages of the game. 
One could potentially mitigate this minor bias in unregularized and decoupled blocks in the policy matrices through heuristic adjustments. 
However, since the impact of the accumulation is minimal in our experiments (up to only 0.35\%), we did not implement specific corrective measures.

\end{appendixsec}

\begin{IEEEbiography}[{\includegraphics[width=1in,height=1.25in,clip,keepaspectratio]{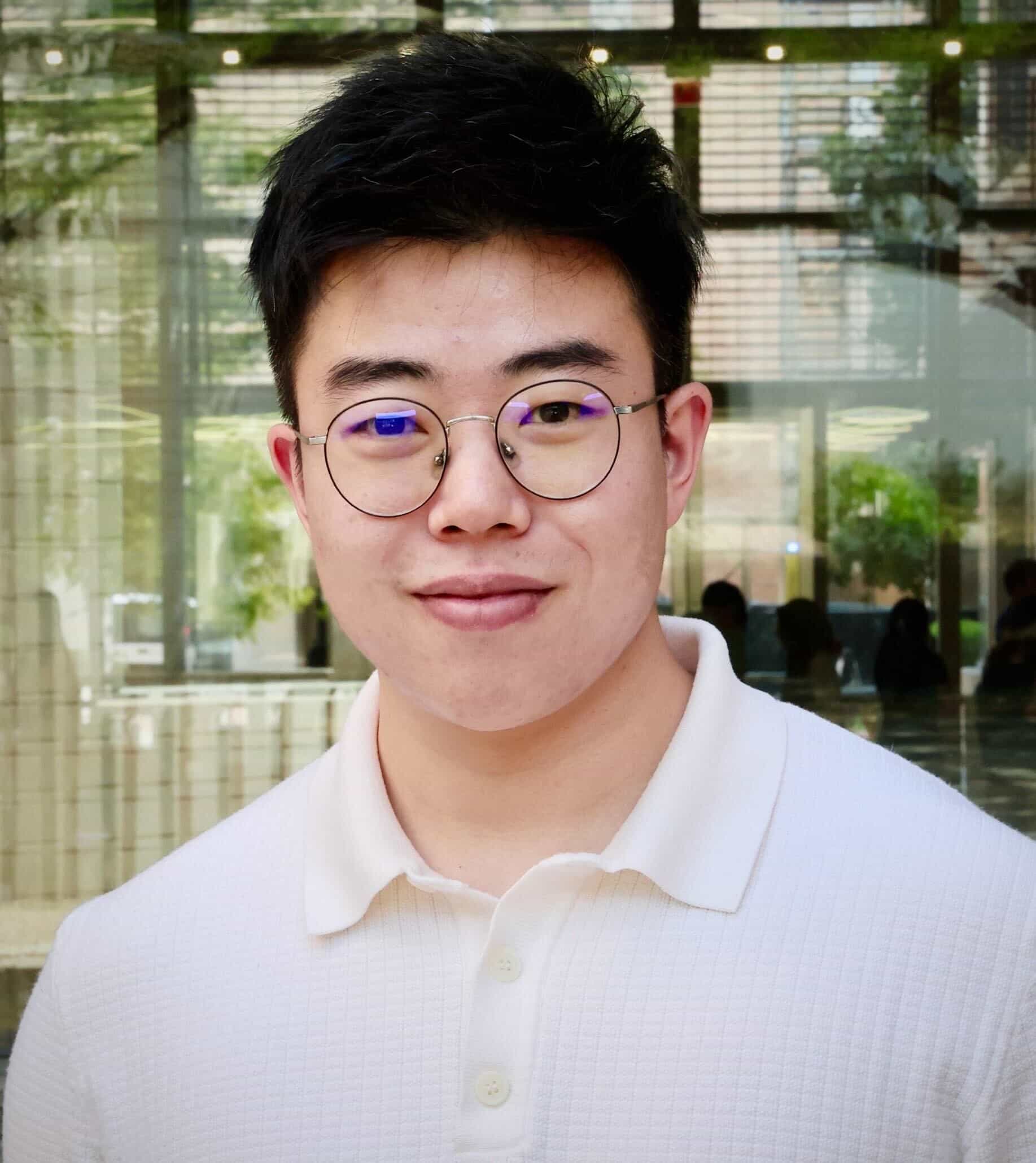}}]{Xinjie Liu} (Graduate Student Member, IEEE) is a Ph.D.\ student in the Department of Electrical and Computer Engineering at The University of Texas at Austin. He received his M.Sc.\ degree (Cum Laude) in Robotics from Delft University of Technology and his B.Eng. degree in Vehicle Engineering (Automobile) from Tongji University. 
His research interests lie in theoretical foundations and practical decision-making and control algorithms for autonomous systems. 
His research often draws on tools from dynamic game theory, reinforcement learning, and optimization.
\end{IEEEbiography}

\begin{IEEEbiography}[{\includegraphics[width=1in,height=1.25in,clip,keepaspectratio]{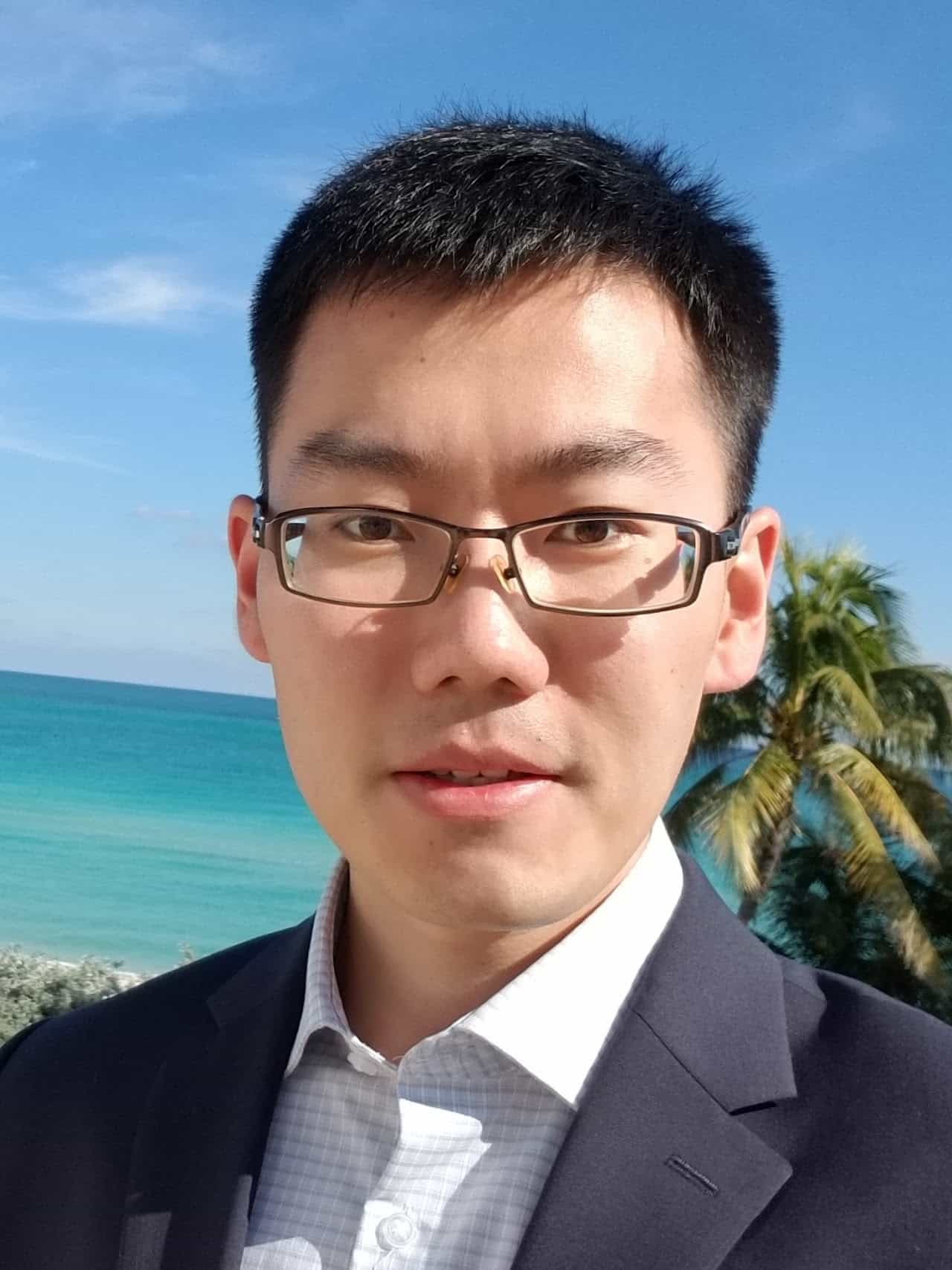}}]{Jingqi Li} (Graduate Student Member, IEEE) 
is a Ph.D.\ candidate in the Department of Electrical Engineering and Computer Sciences at UC Berkeley. He earned his M.S.\ degree in Electrical Engineering from the University of Pennsylvania, where his research was awarded an Outstanding Research Award. Prior to that, he received a B.E.\ degree in Aerospace Engineering from Beijing University of Aeronautics and Astronautics. His research interests center around dynamic game theory, control, deep reinforcement learning, with a focus on enhancing safety and efficiency in multi-agent strategic decision-making for human-centered robotics.
\end{IEEEbiography}

\begin{IEEEbiography}[{\includegraphics[width=1in,height=1.25in,clip,keepaspectratio]{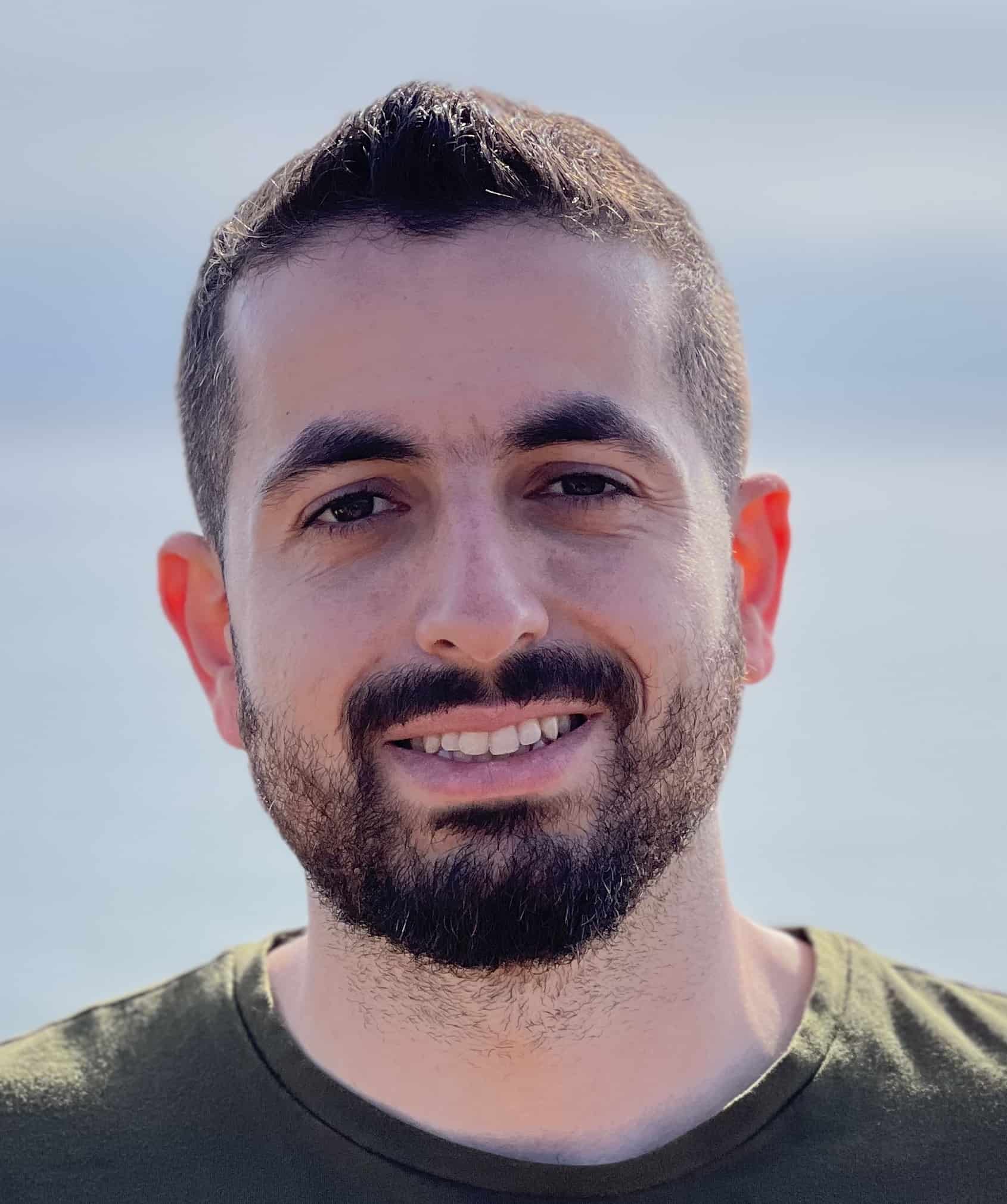}}]{Filippos Fotiadis}
(Member, IEEE) was born in Thessaloniki, Greece. He received the PhD degree in Aerospace Engineering in 2024, and the MS degrees in Aerospace Engineering and Mathematics in 2022 and 2023, all from Georgia Tech. Prior to his graduate studies, he received a diploma in Electrical \& Computer Engineering from the Aristotle University of Thessaloniki. He is currently a postdoctoral researcher at the Oden Institute for Computational Engineering \& Sciences at the University of Texas at Austin.
His research interests are in the intersection of systems \& control theory, game theory, and learning, with applications to the security and resilience of cyber-physical systems.
\end{IEEEbiography}

\begin{IEEEbiography}[{\includegraphics[width=1in,height=1.25in,clip,keepaspectratio]{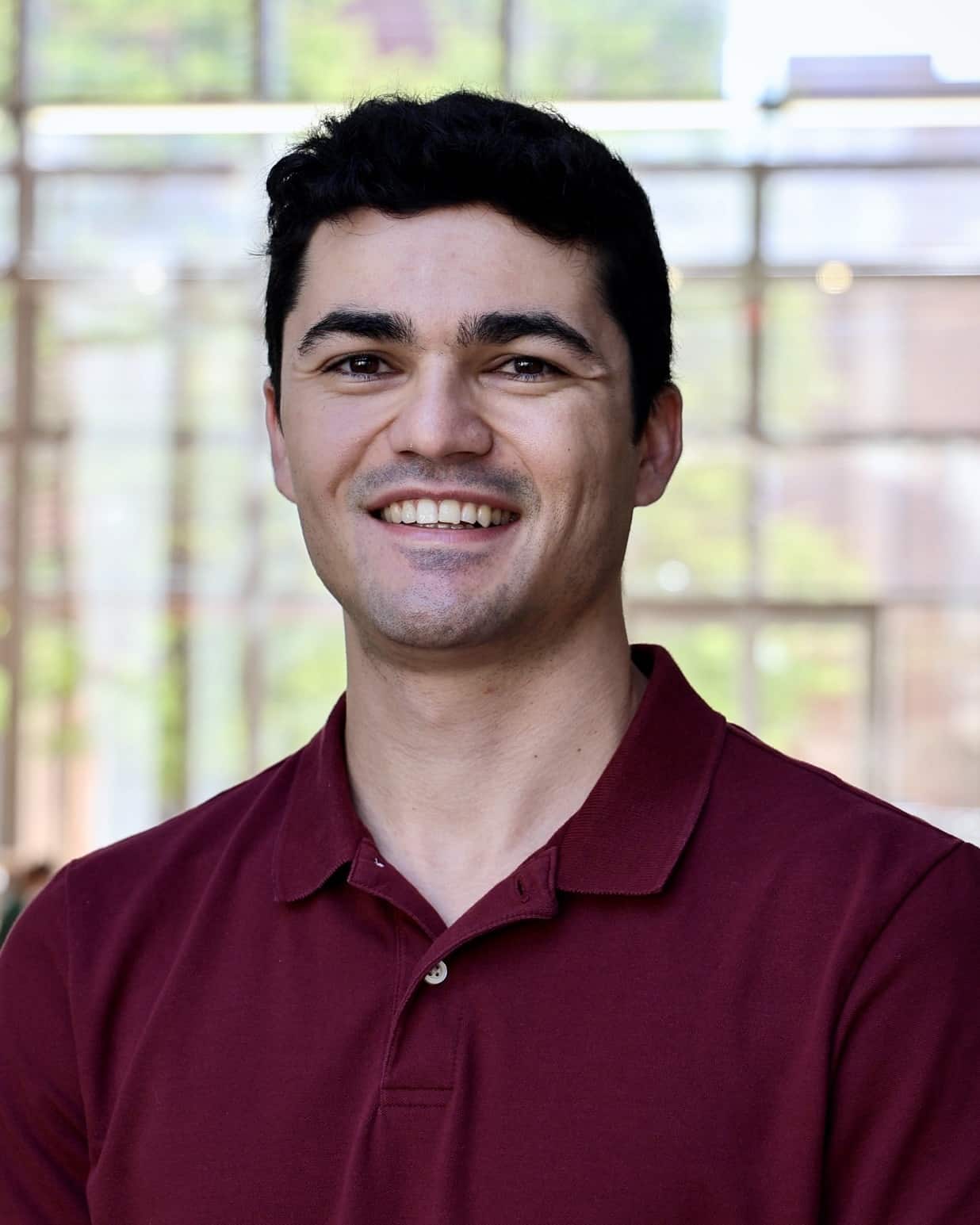}}]{Mustafa O. Karabag} is a postdoctoral fellow in the Oden Institute for Computational Engineering \& Sciences at the University of Texas at Austin. He received his Ph.D. degree from the University of Texas at Austin in 2023. His research focuses on developing theory and algorithms to control the information flow of autonomous systems to succeed in information-scarce or adversarial environments.
\end{IEEEbiography}

\begin{IEEEbiography}[{\includegraphics[width=1in,height=1.25in,clip,keepaspectratio]{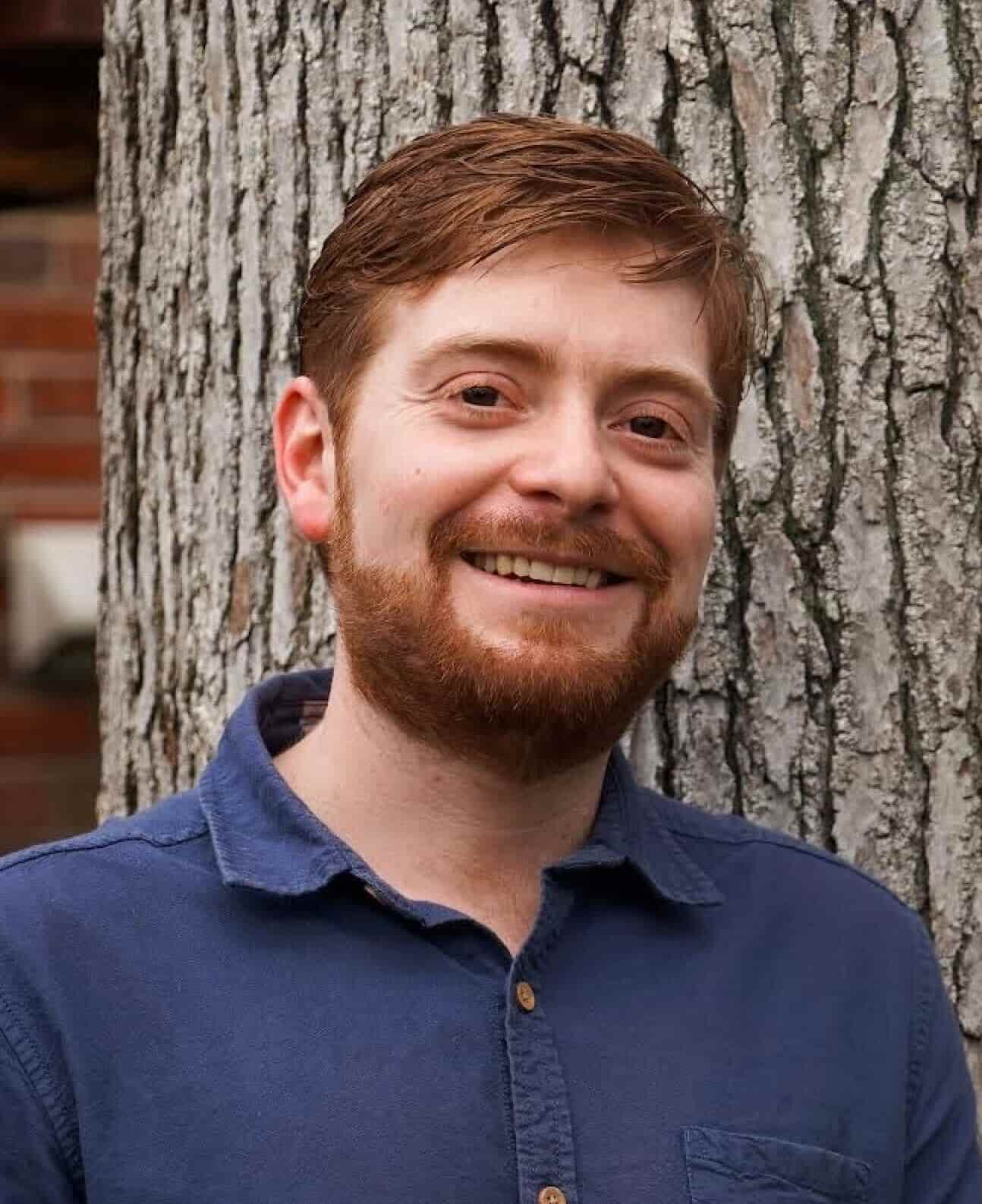}}]{Jesse Milzman} is a research scientist at the Army Research Laboratory and a visiting researcher at New York University, working out of Brooklyn, New York. His main research interests concern the value of information in multi-agent systems, utilizing tools from game theory and information theory. Topics include strategic information gathering, deception, and resilient multi-agent autonomy.  He received his B.S. from Georgetown University (2015), and continued on to receive his Ph.D. from the University of Maryland (2021), both in Mathematics.
\end{IEEEbiography}

\begin{IEEEbiography}[{\includegraphics[width=1in,height=1.25in,clip,keepaspectratio]{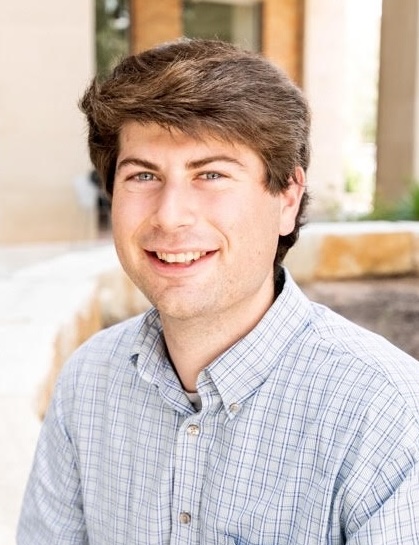}}]{David Fridovich-Keil} (Member, IEEE) received the B.S.E. degree in electrical engineering from Princeton University, and the Ph.D. Degree from the University of California, Berkeley. He is an Assistant Professor in the Department of Aerospace Engineering and Engineering Mechanics at the University of Texas at Austin. Fridovich-Keil is the recipient of an NSF Graduate Research Fellowship and an NSF CAREER Award.
\end{IEEEbiography}


\begin{IEEEbiography}[{\includegraphics[width=1in,height=1.25in,clip,keepaspectratio]{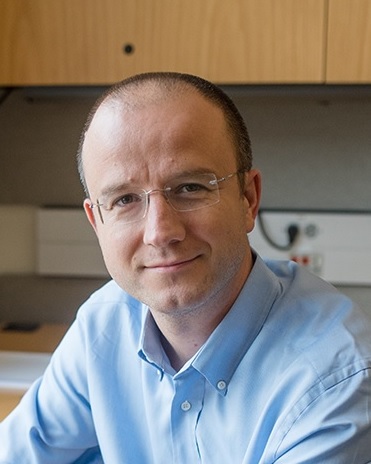}}]{Ufuk Topcu}(Fellow, IEEE)  is currently a Professor
with the Department of Aerospace Engineering and
Engineering Mechanics, The University of Texas at
Austin, Austin, TX, USA, where he holds the Temple
Foundation Endowed Professorship No. 1 Professorship. He is a core Faculty Member with the Oden Institute for Computational Engineering and Sciences and
Texas Robotics and the director of the Autonomous
Systems Group. His research interests include the
theoretical and algorithmic aspects of the design and
verification of autonomous systems, typically in the
intersection of formal methods, reinforcement learning, and control theory.
\end{IEEEbiography}


\end{document}